\documentclass[conference]{IEEEtran}
\IEEEoverridecommandlockouts
\usepackage{cite}
\usepackage{amsmath,amssymb,amsfonts}
\usepackage{algorithmic}
\usepackage{graphicx}
\usepackage{textcomp}
\usepackage{xcolor}
\usepackage{array}
\usepackage{booktabs}
\usepackage{tabularx}
\usepackage{longtable}
\usepackage{placeins}
\usepackage{ragged2e}
\usepackage[hyphens]{url}

\def\BibTeX{{\rm B\kern-.05em{\sc i\kern-.025em b}\kern-.08em
    T\kern-.1667em\lower.7ex\hbox{E}\kern-.125emX}}
\begin{document}

% Ensure letter paper
\pdfpagewidth=8.5in
\pdfpageheight=11in

%%%%%%%%%%%---SETME-----%%%%%%%%%%%%%
\newcommand{\iscasubmissionnumber}{NaN}
\newcommand{\Shifan}[1]{\textcolor{orange}{[Shifan: #1]}}
%%%%%%%%%%%%%%%%%%%%%%%%%%%%%%%%%%%%

\pagenumbering{arabic}

%%%%%%%%%%%---SETME-----%%%%%%%%%%%%%
\title{Identifying Logical Operations from Syndrome Histories for qLDPC Codes}

\author{
\IEEEauthorblockN{Yizhuo Tan}
\IEEEauthorblockA{
Yale University\\
New Haven, CT, USA\\
yizhuo.tan@yale.edu
}
\and
\IEEEauthorblockN{Shifan Xu}
\IEEEauthorblockA{
Yale University\\
New Haven, CT, USA\\
shifan.xu@yale.edu
}
\and
\IEEEauthorblockN{Yongshan Ding}
\IEEEauthorblockA{
Yale University\\
New Haven, CT, USA\\
yongshan.ding@yale.edu
}
\and
\IEEEauthorblockN{Jakub Szefer}
\IEEEauthorblockA{
Northwestern University\\
Evanston, IL, USA\\
jakub.szefer@northwestern.edu
}
}
%%%%%%%%%%%%%%%%%%%%%%%%%%%%%%%%%%%%

\maketitle
\thispagestyle{plain}
\pagestyle{plain}

%%%%%% -- PAPER CONTENT STARTS-- %%%%%%%%

\begin{abstract}
\label{abstract}
Syndrome histories are normally used to detect and correct errors in fault-tolerant quantum computation, but they can also reveal information about the logical operations executed during computation. We study logical-operation identification using syndrome histories alone or together with other classical measurement outcomes. We develop an exact inference method and a scalable approximation that account for unknown circuit faults and ideal measurement randomness.
We evaluate automorphisms, logical Pauli measurements, T injections, and logical circuits combining these instructions on \([[18,4,2]]\) and \([[72,12,6]]\) bivariate bicycle (BB) codes under multiple noise models and error rates.
A single logical instruction can be identified with over 94\% accuracy under circuit-level noise with error rates between \(10^ {-5} \) and \(10^{-3}\). We also achieve over 99\% accuracy for automorphism-based logical circuits and 78.32\% for logical circuits combining all three instructions.
We further analyze identification ambiguities arising from insufficient or indistinguishable measurement information, fault timing, and approximation inference. 
Together, these results extend the role of syndromes beyond error decoding to identifying executed logical operations and circuits during fault-tolerant quantum computation. They also suggest that access to syndrome histories and related classical measurement outcomes may unintentionally reveal information about the executed computation.
\end{abstract}

\section{Introduction}
\label{sec:introduction}

Fault-tolerant quantum computation relies on repeated syndrome extraction during logical circuit execution to detect errors and maintain the correctness of computation~\cite{terhal2015quantum,steane1997active}. The resulting syndrome histories are normally processed by decoders to infer errors and determine corrections~\cite{terhal2015quantum}. In this work, we ask whether the same syndrome histories alone or together with auxillary-qubit measurement outcomes can also reveal which logical operations and logical circuits were executed. We investigate this question for quantum low-density parity-check (qLDPC) codes~\cite{tillich2013quantum,panteleev2022asymptotically} and demonstrate our identification method on bivariate bicycle (BB) codes, which offer a promising approach to reducing the overhead of fault-tolerant quantum computation~\cite{bravyi2005universal,yoder2025tour}.

Universal fault-tolerant quantum computation requires both Clifford and non-Clifford operations~\cite{boykin2000new,eastin2009restrictions}. For BB codes, code automorphisms and logical Pauli measurements with classical feedforward support Clifford operations~\cite{yoder2025tour,cross2024improved}.
Non-Clifford \(T\) operations can be implemented through magic-state injection, which uses a prepared magic state, joint logical measurements, and conditional corrections.~\cite{bravyi2005universal,gottesman1999demonstrating}. 
% \Shifan{split into two sentences. xxx implete clifford. xxx implement non-clifford (expend the t injection a little more ,like t magic state + clifford injection gives non-clifford).} 
We therefore study identification of these three logical instructions, namely code automorphisms, logical Pauli measurements, and \(T\) injections, both individually and as parts of logical circuits. We refer to this task as logical-operation identification.

\begin{figure}[t]
    \centering
    \includegraphics[width=\linewidth]{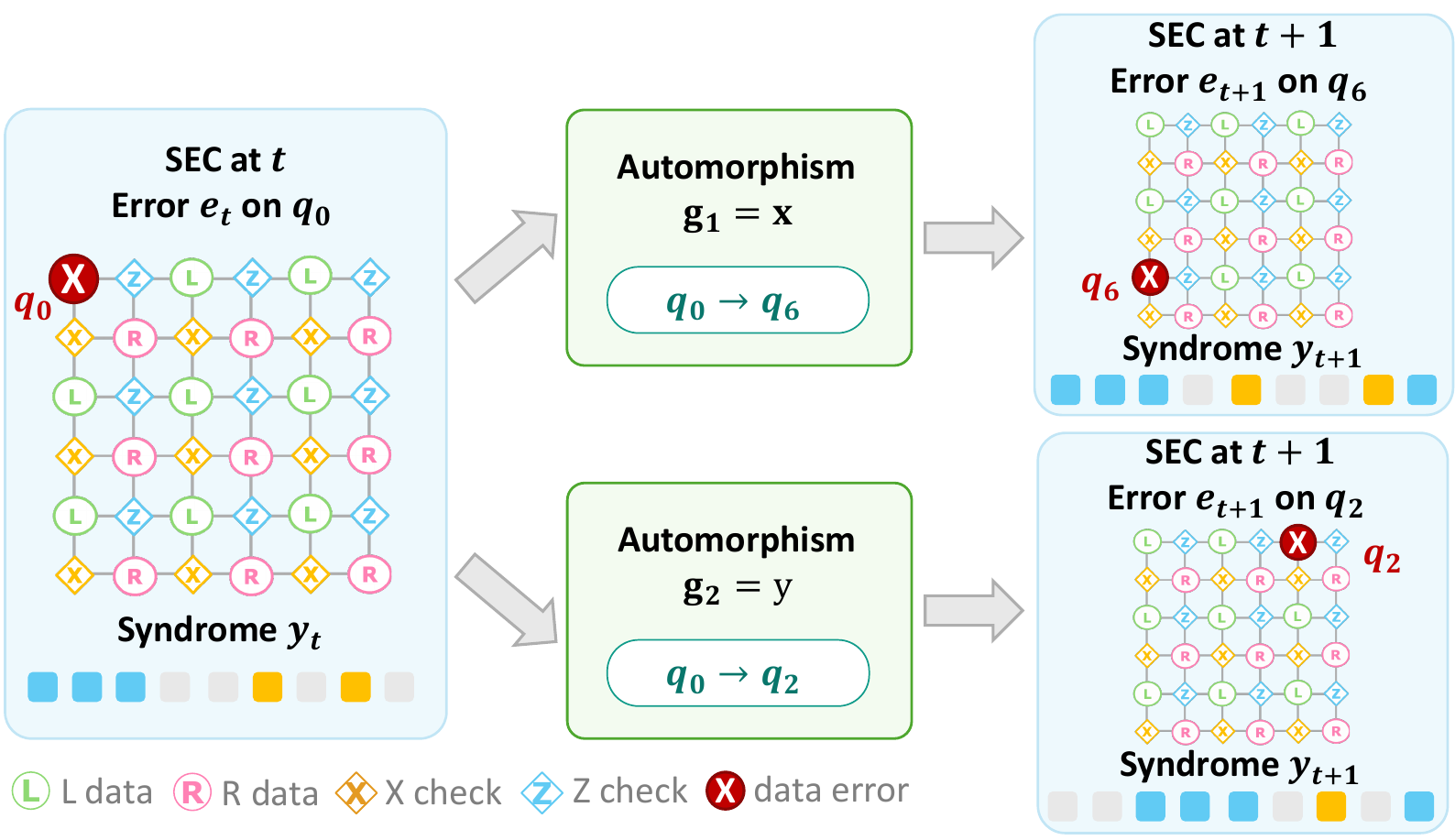}
    \caption{Error transport can make candidate automorphisms distinguishable. The same detectable Pauli error $e_t$ is present before either candidate automorphism. The two automorphisms transport the error to different data qubits, producing different syndromes at the next SEC.
    }
    \label{fig:error_transport}
\end{figure}

A code automorphism is a symmetry of the code that preserves its codespace while permuting the physical qubits~\cite{gottesman1998theory,sayginel2025fault}. This permutation can induce a nontrivial logical operation on the encoded qubits. A data error present before an automorphism is transported according to its physical-qubit permutation. Notably, different automorphisms can transport the same single data qubit error to different qubit locations. A subsequent syndrome-extraction circuit (SEC) can capture this difference because the relocated errors produce different syndromes. Figure~\ref{fig:error_transport} illustrates this mechanism for two candidate automorphisms. 

Logical Pauli measurements and $T$ injections provide another source of identifying information. Their implementations involve additional auxiliary qubits and measurements~\cite{williamson2026low,yoder2025tour}, producing classical measurement outcomes beyond the BB's memory qubit syndrome history. These stabilizer and auxiliary measurement outcomes can have operation-dependent correlations, allowing different instructions to be distinguished even without faults.

% \Shifan{a high level overview of your scheme might be useful here. for example. you first introduce your methods is combining automorphism identification with pauli measurement identification, then you introduce the schemes one by one. think about a better termininology, as your logical-operation identification are used for both: automorphism-identification, pauli measreuement idenfication, and logical circuit identification, which are overusing these words and cause confusing.for example, the following sentence should be any clifford gates can be identified by a combination of automorphism identification and a pauli-measurement identification, something like this. }

We use these two sources of distinguishing information to construct candidate likelihoods for individual logical operations and logical circuits. For logical circuits, the likelihood calculation accounts for error propagation and measurement correlations throughout the circuit, rather than treating each instruction as an independent identification problem.

\begin{figure}[t]
    \centering
    \includegraphics[width=\linewidth]{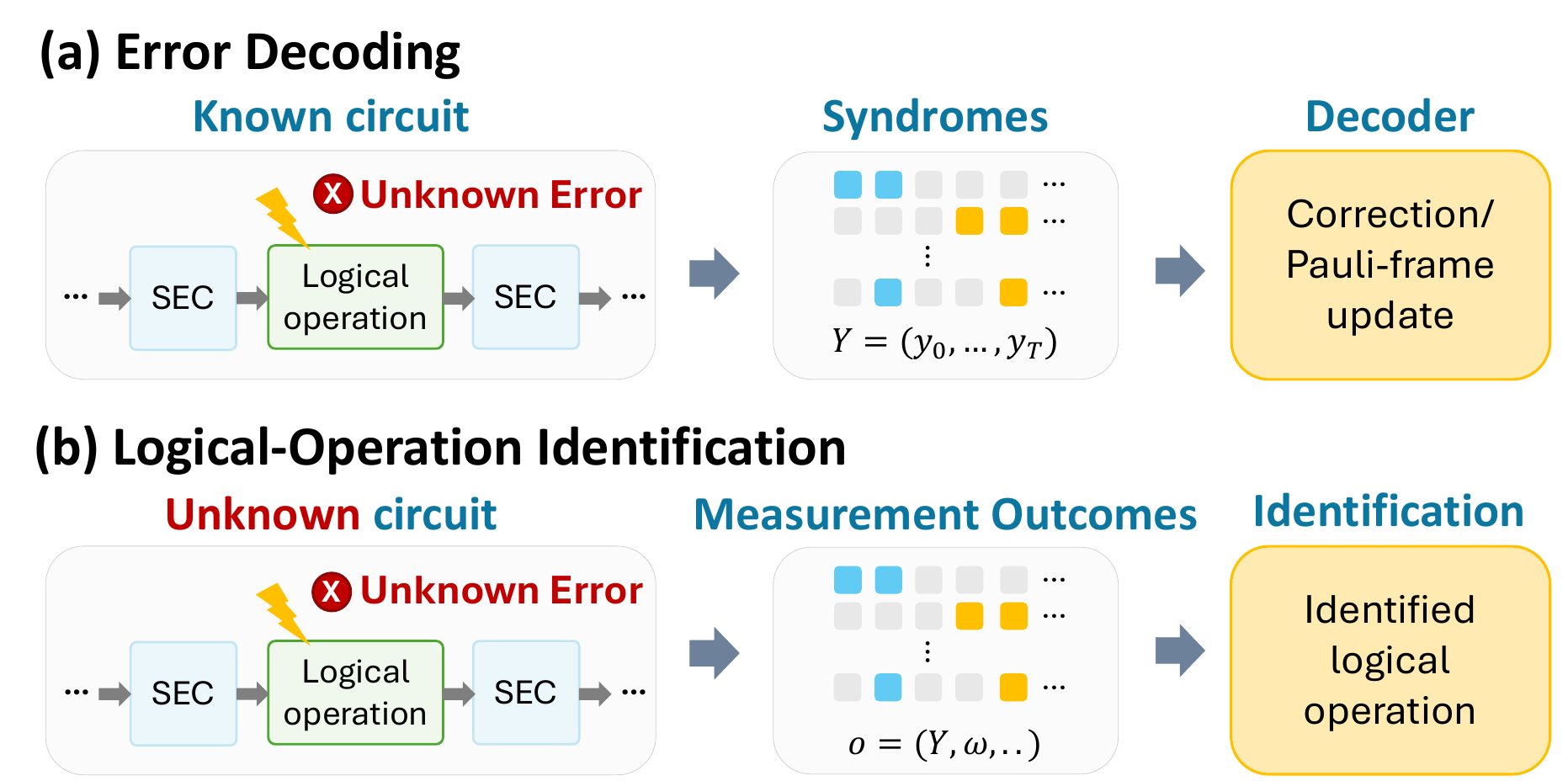}
    \caption{Comparison of error decoding and logical-operation identification. (a) Error decoding assumes the executed circuit is known and uses the syndrome history to determine a correction or Pauli-frame update for unknown errors. (b) Logical-operation identification assumes both the executed circuit and errors are unknown, and uses the same syndrome history, optionally supplemented by auxiliary measurement outcomes, to identify the logical operations.}
    \label{fig:decoding_vs_identification}
\end{figure}

Logical-operation identification therefore uses classical measurement outcomes as error decoding does, but differs in what is known and what is inferred, as illustrated in Figure~\ref{fig:decoding_vs_identification}. In error decoding, the executed circuit is known, and the syndrome history and relevant auxiliary measurement outcomes are used to infer unknown errors and determine a correction or Pauli-frame update~\cite{terhal2015quantum}. In logical-operation identification, the executed logical operation or circuit is unknown and is inferred from the syndrome history, optionally together with additional classical measurement outcomes generated during circuit execution.

However, the classical measurement outcomes do not always contain enough information to identify different logical operations or circuits. Different candidate operations or circuits can produce identical or overlapping measurement outcomes, making some executions ambiguous. Even when the candidates are distinguishable in principle, unknown faults and measurement randomness can produce many possible explanations for the same outcomes. Accounting for these possibilities becomes increasingly expensive for larger codes and longer circuits, leading to substantial computation and memory requirements.

To address these challenges, we introduce exact identification methods and scalable approximations to compute candidate likelihoods for automorphisms, logical Pauli measurements, \(T\) injections, and logical circuits combining these operations.
% \Shifan{THIS IS FOR automorphism or for pauli measurements? be explicit}
The exact methods evaluate candidate likelihoods while accounting for unknown faults and measurement randomness, and combine candidates that share the requested recovery target before making a decision. Scalable inference methods use approximations to reduce the computation required for larger codes and longer circuits.

We evaluate automorphisms, logical Pauli measurements, $T$-injections, and logical circuits combining all three instruction types on \([[18,4,2]]\) and \([[72,12,6]]\) under multiple noise models. The results demonstrate high identification accuracy for logical operations and logical circuits and show that stabilizer syndrome histories alone can support accurate identification. We also analyze the reasons and the corresponding probability for identification failures, including syndrome ambiguity, fault timing, and scalable inference approximation.

The main contributions of this work are as follows.
\begin{itemize}
    \item \textbf{Logical-operation identification from syndrome histories.}
    We formulate logical-operation identification in qLDPC codes and show that operation-dependent error transport and measurement correlations can provide distinguishing information about executed logical operations and circuits in syndrome histories, while auxiliary measurement outcomes can provide additional information.
    \item \textbf{Exact and scalable identification methods.}
    We develop exact methods for accurate logical-operation identification and scalable methods that reduce computation and memory cost.
    \item \textbf{Evaluation across codes, circuits, and noise settings.}
    We evaluate identification of automorphisms, logical Pauli measurements, T injections, and logical circuits on \([[18,4,2]]\) and \([[72,12,6]]\) under different noise models and syndrome-extraction settings.
    \item \textbf{Analysis of identification limitations and refinements.}
    We analyze identification limitations arising from insufficient measurement information, fault timing, and scalable inference approximation. We further evaluate refinements to measurement settings and scalable inference that improve identification accuracy.
\end{itemize}

The remainder of this paper is organized as follows. Section~\ref{sec:background} introduces backgrounds for BB codes, syndrome extraction, and logical instructions supporting universal BB computation. Section~\ref{sec:method} presents our exact and scalable methods for logical-operation identification.
Section~\ref{sec:experimental-setup} describes the experimental setup, and Section~\ref{sec:evaluation} evaluates identification accuracy across logical instructions and circuits.
Section~\ref{sec:limit} analyzes identifiability, inference limits and refinements. Related works are discussed in Section~\ref{sec:related_work}, with Section~\ref{sec:discussion} presents discussion and future directions. Finally, Section~\ref{sec:conclusion} concludes our work.

\section{Background}
\label{sec:background}

\subsection{Stabilizer Codes, BB Codes, and Syndrome Extraction}
\label{subsec:background-codes}

An \([[n,k,d]]\) stabilizer code encodes \(k\) logical qubits into \(n\) physical qubits and has code distance \(d\), defined as the minimum weight of a nontrivial logical Pauli operator~\cite{gottesman1997stabilizer}. Its stabilizer group \(\mathcal S\) is generated by commuting Pauli checks whose measurements detect data errors. A Calderbank-Shor-Steane (CSS) code uses separate \(X\)-type and \(Z\)-type stabilizer generators, represented by binary matrices \(H_X\) and \(H_Z\) satisfying \(H_XH_Z^{\mathsf T}=0\)~\cite{calderbank1996good,steane1996error}. Quantum low-density parity-check (qLDPC) codes have sparse stabilizer checks, with bounded check weights and qubit participation as the code size increases~\cite{tillich2013quantum,panteleev2022asymptotically}. Bivariate bicycle (BB) codes are a family of CSS qLDPC codes constructed from two commuting sparse matrices \(A\) and \(B\), defined by bivariate polynomials in cyclic-shift matrices~\cite{bravyi2005universal}. Their stabilizer check matrices take the form
\[
H_X= \begin{bmatrix} A&B \end{bmatrix}, \qquad H_Z= \begin{bmatrix} B^{\mathsf T} & A^{\mathsf T} \end{bmatrix}.
\]
Ignoring scalar phases, a data error is represented by a binary symplectic vector \(e=(e_X, e_Z)^{\mathsf T}\in\mathbb F_2^{2n}\), where \(e_X\) and \(e_Z\) specify its \(X\) and \(Z\) components. Ordering the outcomes of the \(Z\)-type checks before those of the \(X\)-type checks, the corresponding data syndrome is
\[
s=Je, \qquad J= \begin{bmatrix} H_Z&0\\ 0&H_X \end{bmatrix}.
\]
All binary operations are over \(\mathbb F_2\). Data errors with the same syndrome need not have the same logical effect because a nontrivial logical Pauli operator can commute with every stabilizer while acting nontrivially on the encoded state. In particular, a zero data syndrome does not exclude a nontrivial logical Pauli error.

A syndrome-extraction circuit (SEC) measures the stabilizer generators using ancillary qubits coupled to the data~\cite{steane1997active,bravyi2024high}. Repeated SEC rounds produce a syndrome history \(Y=(y_0,\ldots,y_T)\), where \(y_t\) denotes the syndrome at boundary \(t\). Under a measurement-error model,
\[
y_t=s_t \oplus m_t,
\]
where \(m_t\) denotes the binary vector of measurement errors affecting the recorded stabilizer outcomes. More generally, faults during preparation, gates, idling, and measurement can affect both the data errors and subsequent measurement outcomes. Syndrome extraction provides information about these errors but does not itself remove them. Using the syndrome history, an error decoder may determine a physical Pauli correction or a Pauli-frame update, which tracks the correction in software and adjusts the interpretation of later operations and measurements~\cite{terhal2015quantum}.

\subsection{Logical Instructions for Universal BB Computation}
\label{subsec:background-instructions}

Universal quantum computation can be implemented using the Clifford+\(T\) gate set~\cite{boykin2000new}. Clifford operations map Pauli operators to Pauli operators under conjugation, while the \(T\) gate provides a non-Clifford operation needed for universality~\cite{gottesman1997stabilizer}. Architectures based on BB codes combine code automorphisms, logical Pauli measurements, and T injection to support universal logical computation~\cite{yoder2025tour}.

Code automorphisms implement logical Clifford gates through code-preserving permutations of physical qubits~\cite{sayginel2025fault}. Logical Pauli measurements project the encoded state onto an eigenspace of a logical Pauli operator. They can be implemented through gauging, which couples data qubits to auxiliary qubits in a logical processing unit (LPU) and measures additional commuting check operators to obtain the logical Pauli eigenvalue~\cite{williamson2026low,yoder2025tour}. Subsequent ungauging returns the system to the original code. Joint Pauli measurements between BB code blocks can be implemented using code-code adapters~\cite{yoder2025tour,cross2024improved}. T injection implements non-Clifford Pauli rotations using a prepared magic state and a joint logical Pauli measurement. Measurement outcomes may require physical corrections or classical frame updates that affect subsequent operations~\cite{gottesman1999demonstrating,terhal2015quantum}. During execution, SEC rounds provide measured syndromes, while gauging and T injection also generate auxiliary-qubit and factory measurement outcomes~\cite{yoder2025tour}. These classical outcomes can provide information for identifying the executed logical instructions.

\subsection{Code Automorphisms and Logical Actions}
\label{sec:background-automorphisms}

A qubit-permutation automorphism \(g\) permutes the physical data qubits while preserving the stabilizer code~\cite{gottesman1998theory,sayginel2025fault}. Let \(P_g\) denote the permutation matrix of \(g\). The corresponding unitary \(U_g\) satisfies \(U_g\mathcal S U_g^\dagger=\mathcal S,\) so the permutation preserves the codespace and may induce a nontrivial logical gate. BB codes admit cyclic-shift automorphisms, while additional code-preserving permutations can occur for particular code instances~\cite{bravyi2024high,sayginel2025fault}. The action of an automorphism \(g\) is characterized at three levels. Its physical symplectic action \(S_g\) describes how data errors transform, its syndrome action \(R_g\) describes how the corresponding data syndromes transform, and its logical action \(\Lambda_g\) describes the induced transformation of encoded Pauli operators~\cite{gottesman1997stabilizer,sayginel2025fault}. For a qubit permutation,
\[
e'=S_ge, \qquad S_g=\operatorname{diag}(P_g,P_g).
\]
Let \(L\in\mathbb F_2^{2n\times 2k}\) contain binary representatives of a chosen logical Pauli basis \((\bar X_1,\ldots,\bar X_k,\bar Z_1,\ldots,\bar Z_k)\). The three actions satisfy
\[
JS_g=R_gJ, \qquad S_gL\equiv_{\mathcal S}L\Lambda_g,
\]
where \(\equiv_{\mathcal S}\) denotes equivalence up to stabilizers.
The first relation gives \(s'=R_gs\) for ideal error transport. The second gives the logical action modulo stabilizers. These representations characterize the effects of the same automorphism on data errors, data syndromes, and encoded Pauli operators. The logical action \(\Lambda_g\) specifies the transformation of encoded Pauli operators up to phases, rather than the complete logical gate.

An ordered automorphism sequence \(h=(g_1,\ldots,g_q)\) contains \(q\) automorphisms in chronological order. We denote by \(\mathcal H\) the candidate set of ordered automorphism sequences considered by the identification procedure and refer to \(h\in\mathcal H\) as a candidate sequence. Its net physical action and net logical action are
\[
S_h = S_{g_q}\cdots S_{g_1}, \qquad \Lambda_h = \Lambda_{g_q}\cdots\Lambda_{g_1}.
\]
For an automorphism sequence, \(\Gamma(h)\) denotes the logical circuit induced by \(g_1,\ldots,g_q\). Different ordered automorphism sequences may induce the same logical circuit, and different logical circuits may have the same net logical action.

\subsection{Gauging-Based Logical Pauli Measurements}
\label{subsec:background-pm}
A logical Pauli measurement instruction \(M_P\) measures a logical Pauli operator \(P\), with physical Pauli representative \(\widetilde P\) on the data qubits. Its binary outcome \(\mu\) labels the eigenvalue \((-1)^\mu\) and the corresponding projector \(\Pi_\mu^{(P)}=(I+(-1)^\mu P)/2\). Gauging implements this measurement through local checks involving auxiliary qubits instead of measuring the high-weight operator \(\widetilde P\) directly~\cite{williamson2026low}.

Gauging temporarily extends the code with auxiliary edge qubits on a graph \(G=(V,E)\). The representative \(\widetilde P\) is divided into local Pauli factors \(L_v^{(P)}\) at the vertices. The Gauss's law operators satisfy
\[
A_v^{(P)}=L_v^{(P)}\prod_{e\ni v}X_e, \qquad \prod_{v\in V}A_v^{(P)}=\widetilde P,
\]
under the fixed Pauli sign convention. Thus, the parity of the Gauss measurement outcomes determines \(\mu\). Additional checks are needed to maintain consistent stabilizer constraints in the temporary gauged code.

Flux operators enforce constraints around cycles of the auxiliary graph through products of edge-\(Z\) operators~\cite{williamson2026low}. To maintain compatibility with the Gauss's law operators, original BB stabilizers are modified into deformed stabilizer checks,
\[
\widetilde S_j^{(P)} = S_j\prod_{e\in E_j^{(P)}}Z_e,
\]
where \(E_j^{(P)}\) is chosen so that \(\widetilde S_j^{(P)}\) commutes with the Gauss's law operators. These modified checks preserve stabilizer information during gauging and relate it to the auxiliary edge qubits.

After the logical measurement, ungauging measures the auxiliary edge qubits in the \(Z\) basis and restores the original BB stabilizer structure up to measurement-dependent Pauli byproducts~\cite{williamson2026low}. These byproducts can be corrected physically or tracked in a Pauli frame.

In the two-round gauging schedule considered in this work, the classical measurement outcomes include the BB syndrome measured before gauging and after ungauging, two rounds of Gauss, flux, and deformed stabilizer outcomes, and the auxiliary edge-\(Z\) readouts collected during ungauging.

\subsection{T Injection}
\label{subsec:background-t}
Non-Clifford \(T\) operations are commonly implemented using logical \(T\) states prepared separately in magic-state factories~\cite{bravyi2005universal,yoder2025tour}. These states are then injected into the computation through logical Pauli measurements and conditional corrections. For a logical Pauli operator \(P\), the T-injection instruction \(T_P\) implements the rotation \(U_P=e^{i\pi P/8}\) using a prepared magic state \(|T\rangle=e^{i\pi Z_T/8}|+\rangle,\) where \(X_T\) and \(Z_T\) are the logical Pauli operators acting on the encoded resource state. For \(P=Z\), this positive-exponent rotation corresponds to the conventional \(T^\dagger\) gate up to a global phase.

The injection begins with the joint logical Pauli measurement \(M_{Z_T\otimes P}\), which measures \(Z_T\otimes P\) between the prepared logical \(T\) state and the encoded data~\cite{gottesman1999demonstrating,yoder2025tour}. It uses the gauging procedure described above, with one Gauss's law operator extended by \(Z_T\). This produces the binary outcome \(m\), together with the associated gauging and ungauging outcomes. The logical qubit carrying the \(T\) state is then destructively measured in the \(X_T\) basis, producing a second binary outcome \(n\), with \(m,n\in\{0,1\}\)~\cite{yoder2025tour}.

The outcomes \(m,n\) determine the resulting operation on the encoded data. The outcome \(m\) determines the sign of the Pauli rotation, while \(n\) determines a Pauli byproduct. A conditional Clifford correction ensures the intended rotation \(U_P\). Pauli byproducts can be tracked in a Pauli frame, while conditional Clifford corrections can be applied or deferred by adjusting subsequent logical operations and measurement bases~\cite{yoder2025tour}.

\section{Logical-Operation Identification}
\label{sec:method}
The syndrome and auxiliary measurement outcomes produced during logical computation can reveal which logical instructions were executed. Automorphisms transport data errors, potentially changing the syndromes measured by subsequent SEC rounds. Logical Pauli measurements produce candidate-dependent correlations among gauging, ungauging, and stabilizer measurement outcomes, while T injection uses a joint logical Pauli measurement whose classical measurement outcomes can reveal the axis of the injected rotation. We formulate the candidate likelihood for each instruction family and then extend the identification method to logical circuits containing them.

\subsection{Identification Model and Recovery Targets}
\label{subsec:method-model}
For a given qLDPC code, let \(\mathcal H\) denote a finite candidate set of supported logical instructions or ordered instruction sequences. Each candidate \(h\in\mathcal H\) specifies the physical gates, measurements, and prescribed feedforward rules for its execution. The code, candidate set, initial encoded state, compiled physical circuits, noise model, and rules for applying or tracking corrections are known to inference. The executed candidate and the faults occurring during execution are unknown. Let \(o\) denote the classical measurement outcomes available for identification, including the measured syndrome history \(Y\) and any auxiliary or logical-resource measurements used by the candidate circuits.

For candidate \(h\), the candidate likelihood \(P(o\mid h)\) is the probability of obtaining the classical measurement outcomes under its physical circuit and the specified noise model. Let \(F\) denote a complete fault pattern, specifying the modeled faults throughout that execution. The candidate likelihood is
\[
P(o\mid h)=\sum_F P(F\mid h)P(o\mid F,h).
\]
The conditional probability \(P(o\mid F,h)\) includes intrinsic quantum measurement randomness even when no faults occur. For circuits with feedforward, the joint probability accounts for the physical gates and fault locations selected by the preceding measurement outcomes.

The recovery target specifies which aspect of the candidate is to be identified. Let \(\tau:\mathcal H\rightarrow\mathcal C\) assign candidate \(h\) to recovery target \(c=\tau(h)\), where \(\mathcal C\) is the set of possible recovery targets. For automorphism sequences, the target may be the ordered sequence \(h\), its net physical action \(S_h\), its logical circuit \(\Gamma(h)\), or its net logical action \(\Lambda_h\). For a logical Pauli measurement \(M_P\) or T-injection instruction \(T_P\), the target identifies the executed instruction through its logical Pauli operator \(P\). For a logical circuit, the target is \(\Gamma(h)\), including its instruction order and prescribed measurement-dependent operations. Given the candidate prior \(\pi(h)\), the unnormalized target probability and target probability are
\[
\begin{aligned}
Z_c(o)&=\sum_{h:\tau(h)=c}\pi(h)P(o\mid h),\\[4pt]
P(c\mid o)&=\frac{Z_c(o)}{\displaystyle\sum_{c'\in\mathcal C}Z_{c'}(o)}.
\end{aligned}
\]
The maximum-a-posteriori (MAP) estimate selects the recovery target maximizing \(Z_c(o)\). Candidates sharing a recovery target are summed before this decision, so selecting the most likely individual candidate and then projecting its target need not give the same answer.

\subsection{Automorphism Identification}
\label{subsec:method-automorphism}
Automorphism identification compares the probabilities of measured syndrome histories under different candidate automorphism sequences. The inference must account for how each sequence transports data errors and how faults affect subsequent syndrome measurements. We first derive these effects, then present exact inference for computing candidate likelihoods and scalable inference for reducing the required computation. For automorphism identification, the classical measurement outcomes consist of the syndrome history \(Y\). The candidate likelihood and target probabilities are therefore written \(P(Y\mid h)\), \(Z_c(Y)\), and \(P(c\mid Y)\).

\begin{figure}[t]
\centering
\includegraphics[width=\linewidth]{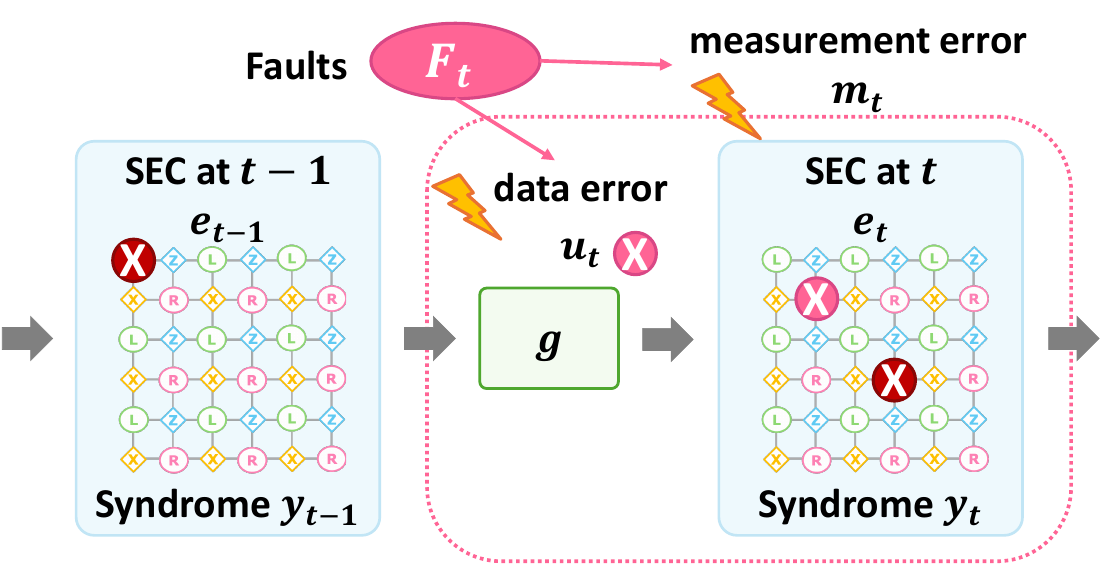}
\caption{Coupling across a noisy syndrome history. The circuit interval contains one automorphism operation $g$ followed by syndrome extraction. Interval fault pattern \(F_t\) determines new data error \(u_t\) and measurement error \(m_t\). Together with the existing data error $e_{t-1}$ transported by $g$, these effects determine the current data error $e_t$ and the measured syndrome $y_t$. Since $e_t$ becomes input to the next operation-SEC interval, adjacent syndromes are coupled and cannot be scored independently.}
\label{fig:noisy_history}
\end{figure}

\subsubsection{Error Transport and Syndrome Change}
\label{subsubsec:auto-error-transport}
We first describe how existing data errors are transported and how new data and measurement errors change the syndrome history. Syndrome boundaries $0,\ldots,T$ divide the circuit into $T$ circuit intervals, each of which may contain one or more automorphisms. For candidate automorphism sequence $h=(g_1,\ldots,g_q)$, let $S_t^{(h)}$ denote the cumulative physical symplectic action of the automorphisms in $h$ executed between syndrome boundaries $t-1$ and $t$, and let $R_t^{(h)}$ denote the corresponding syndrome action. Before any separately modeled correction,
\begin{equation}
    \begin{aligned}
        \mathbf e_t &= S_t^{(h)}\mathbf e_{t-1}\oplus\mathbf u_t,\\
        \mathbf s_t &= R_t^{(h)}\mathbf s_{t-1}\oplus J\mathbf u_t,
    \end{aligned}
    \label{eq:interval_transport}
\end{equation}
where $\mathbf e_t$ is the accumulated data error at boundary $t$, $\mathbf s_t=J\mathbf e_t$ is its data syndrome, and $\mathbf u_t$ is the new data error within the interval after propagation to the outgoing boundary.

Using the measurement-error model from Section~\ref{subsec:background-codes},
\[
    \mathbf y_t=\mathbf s_t\oplus\mathbf m_t,
\]
where $\mathbf m_t$ is the measurement error at boundary $t$. If no correction is applied between the two boundaries,
\[
    \mathbf y_t\oplus R_t^{(h)}\mathbf y_{t-1}
    = J\mathbf u_t\oplus\mathbf m_t\oplus R_t^{(h)}\mathbf m_{t-1}.
\]
We denote by $F_t$ the fault pattern in circuit interval $t$, which specifies the modeled fault events occurring within that interval. For a given incoming state and candidate circuit, the fault pattern determines the new data error \(u_t\) and measurement error \(m_t\) associated with the outgoing syndrome boundary. Under a circuit-level noise model, these faults can occur at different circuit locations and on different qubits throughout the compiled automorphism and syndrome-extraction circuits. Their effects at the outgoing syndrome boundary depend on where they occur and on the subsequent gates through which they propagate.

Figure~\ref{fig:noisy_history} illustrates these effects for an interval containing one automorphism followed by an SEC. The existing data error $\mathbf e_{t-1}$ is transported by the automorphism, while new data error $\mathbf u_t$ and measurement error $\mathbf m_t$ can arise within the interval. Together they determine the outgoing data error $\mathbf e_t$ and measured syndrome $\mathbf y_t$. Because the outgoing data error becomes the input to the next circuit interval, syndrome measurements are coupled through the accumulated data error. The exact calculation must therefore preserve these correlations across syndrome boundaries rather than assign independent probabilities to individual intervals.

\subsubsection{Exact Automorphism Identification}
\label{subsubsec:auto-exact-inference}
For an ordered automorphism sequence $h$, the candidate likelihood $P(Y\mid h)$ defined in Section~\ref{subsec:method-model} can be written as
\begin{equation}
    P(Y\mid h) =
    \sum_{\substack{F:\,F\text{ produces }Y\\\text{under }h}}
    P(F\mid h),
    \label{eq:candidate_likelihood}
\end{equation}
Each complete fault pattern that produces $Y$ enters the sum with its probability $P(F\mid h)$ under the declared noise model. The sum therefore includes different fault locations and fault types, as well as multiple faults and cancellations, rather than retaining only one complete fault pattern. The method evaluates this candidate likelihood one circuit interval at a time through the three steps below.

\begin{figure*}[t]
\centering
\includegraphics[width=\linewidth]{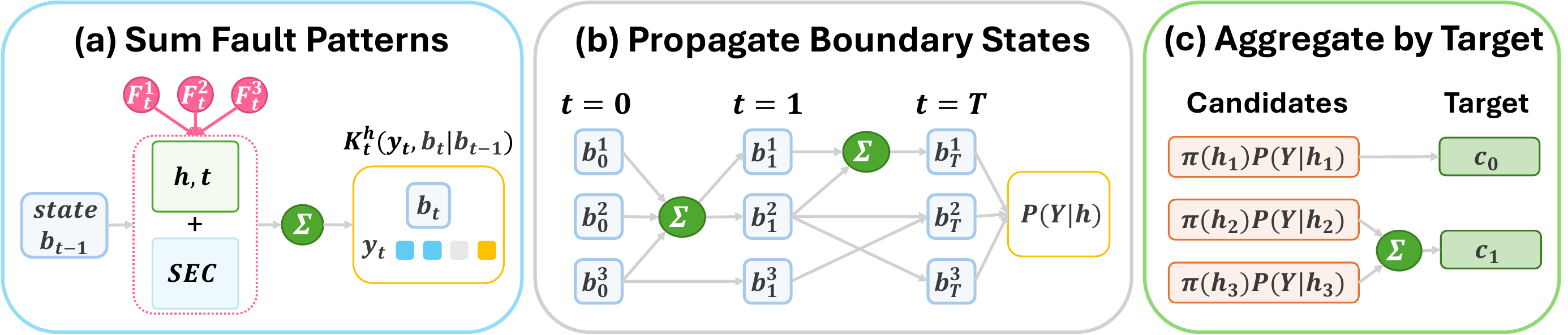}
\caption{Exact automorphism identification.
(a) For candidate sequence $h$ and incoming hidden boundary state $b_{t-1}$, the method sums the probabilities of interval fault patterns $F_t^1,F_t^2,\ldots$ that produce measured syndrome $\mathbf y_t$ and outgoing hidden boundary state $b_t$, yielding the transition probability $K_t^h(\mathbf y_t,b_t\mid b_{t-1})$.
(b) Transition probabilities are composed across syndrome boundaries. Probabilities associated with the same hidden boundary state are summed before the next circuit interval, yielding the candidate likelihood $P(Y\mid h)$.
(c) Prior-weighted candidate likelihoods are summed over candidate sequences that share recovery target $c=\tau(h)$, yielding the unnormalized target probability $Z_c(Y)$ used for the final decision.}
\label{fig:exact_inference}
\end{figure*}

First, the method sums over interval fault patterns to compute the transition probability for each circuit interval.
A hidden boundary state $b_t$ contains the information at syndrome boundary $t$ needed to calculate subsequent measured syndromes for candidate sequence $h$. It contains the data-error state and any additional circuit information that must persist across the boundary.
For candidate sequence $h$, an incoming hidden boundary state $b_{t-1}$, and interval fault pattern $F_t$, the compiled circuit in interval $t$ determines the measured syndrome $\mathbf y_t$ and outgoing hidden boundary state $b_t$. We write
\begin{equation}
    \Psi_t^h(b_{t-1},F_t) = (\mathbf y_t,b_t).
    \label{eq:interval_map}
\end{equation}
The joint syndrome-state transition probability is
\begin{equation}
    K_t^h(\mathbf y_t,b_t\mid b_{t-1})
    = P(\mathbf y_t,b_t\mid b_{t-1},h).
    \label{eq:transition_probability}
\end{equation}
We refer to $K_t^h$ as the transition probability below. It is computed by summing over all interval fault patterns:
\begin{equation}
\begin{aligned}
K_t^h(\mathbf y_t,b_t\mid b_{t-1})
&= \sum_{F_t} P(F_t\mid b_{t-1},h)\\
&\quad\times\mathbf 1\!\left[
\Psi_t^h(b_{t-1},F_t)=(\mathbf y_t,b_t)
\right].
\end{aligned}
\label{eq:fault_pattern_sum}
\end{equation}
The indicator \(\mathsf{1}[\cdot]\) is one when its condition holds and zero otherwise. Figure~\ref{fig:exact_inference}(a) illustrates this step using several possible interval fault patterns $F_t^1,F_t^2,F_t^3$. Their probabilities are added when they produce the same measured syndrome and outgoing hidden boundary state.

Next, the method uses the transition probabilities to propagate forward probabilities across syndrome boundaries. Define the forward probability
\begin{equation}
    \alpha_t(b_t;h)
    = P(b_t,\mathbf y_0,\ldots,\mathbf y_t\mid h).
    \label{eq:forward_probability}
\end{equation}
The forward probability is initialized and updated as
\begin{align}
    \alpha_0(b_0;h) &= P(b_0,\mathbf y_0\mid h),\\
    \alpha_t(b_t;h) &=
    \sum_{b_{t-1}}\alpha_{t-1}(b_{t-1};h)
    K_t^h(\mathbf y_t,b_t\mid b_{t-1}),
    \label{eq:forward_update}
\end{align}
and after the final syndrome boundary,
\begin{equation}
    P(Y\mid h) = \sum_{b_T}\alpha_T(b_T;h).
    \label{eq:terminal_likelihood}
\end{equation}
As shown in Figure~\ref{fig:exact_inference}(b), probabilities associated with the same hidden boundary state are summed before the next circuit interval is processed. Repeating this step across syndrome boundaries yields $P(Y\mid h)$ without explicitly enumerating every complete fault pattern $F$.

Finally, after $P(Y\mid h)$ is computed for every candidate sequence, the candidate prior is applied and candidate likelihoods with the same recovery target are combined using $Z_c(Y)$. Figure~\ref{fig:exact_inference}(c) illustrates this aggregation. Normalization gives $P(c\mid Y)$, and the recovery target with the largest $Z_c(Y)$ is selected.

We refer to this method as Exact-Sum. We also compare it with Exact-Max, which replaces the sum over complete fault patterns for each candidate sequence with the probability of the single most likely complete fault pattern that produces $Y$:
\[
\max_{\substack{F:\,F\text{ produces }Y\\\text{under }h}}
P(F\mid h).
\]
Both methods use the same candidate prior, and candidate sequences are combined by recovery target in the same way.

As the code size and circuit length increase, the number of modeled fault locations also increases, which can rapidly enlarge the set of possible interval fault patterns $F_t$. Uncertainty in the accumulated data error can also increase the number of hidden boundary states propagated across syndrome boundaries, and the transition and forward probabilities must be evaluated for each candidate sequence $h\in\mathcal H$. Although the interval formulation avoids explicitly enumerating every complete fault pattern $F$, the computational cost of exact inference can still grow rapidly with the fault-pattern space, hidden-boundary-state space, and candidate set.

\subsubsection{Scalable Automorphism Identification}
\label{subsubsec:auto-scalable-inference}

The exact identification method becomes expensive when many interval fault patterns must be evaluated within each circuit interval and many hidden boundary states must be propagated across syndrome boundaries. The scalable method presented in this section approximates the candidate likelihood $P(Y\mid h)$ by limiting these two quantities independently, while retaining the same recovery targets and candidate priors as the exact method. We denote the fault-pattern budget by \(B_F\) and the boundary-state budget by \(B_S\), and use a tilde to denote the corresponding approximate quantities used in scalable inference.

\begin{figure*}[t]
\centering
\includegraphics[width=\linewidth]{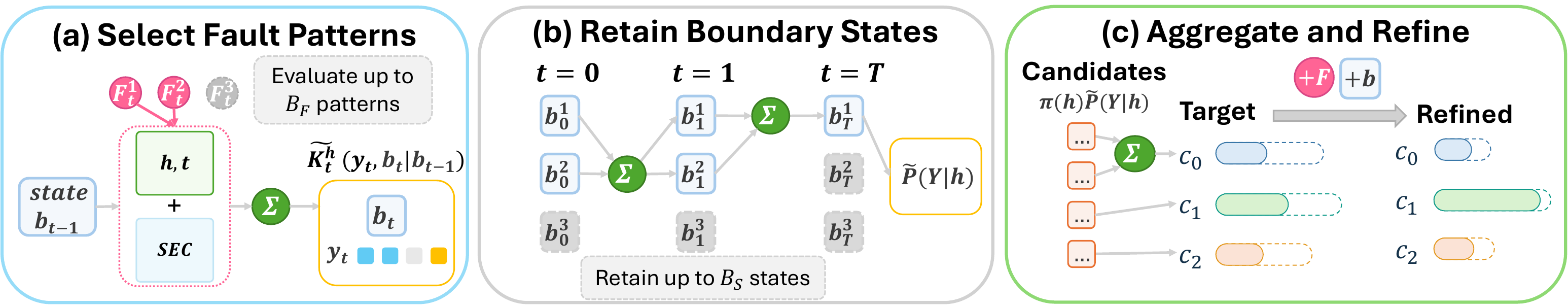}
\caption{Scalable automorphism identification.
The scalable method follows the same three steps as the exact method while limiting the fault-pattern and hidden-boundary-state spaces. (a) For candidate sequence $h$ and incoming hidden boundary state $b_{t-1}$, up to $B_F$ selected interval fault patterns are evaluated to construct the approximate transition probability $\widetilde K_t^h(\mathbf y_t,b_t\mid b_{t-1})$. (b) The resulting probabilities are propagated and summed for equal hidden boundary states, after which up to $B_S$ states are retained for the next circuit interval. Repeating this step gives the approximate candidate likelihood $\widetilde P(Y\mid h)$. (c) Approximate candidate likelihoods are combined over candidate sequences that share the same recovery target, yielding \(\widetilde Z_c(Y)\). When further refinement is needed, additional fault patterns can be evaluated or additional hidden boundary states can be retained.}
\label{fig:scalable_inference}
\end{figure*}

First, the scalable method limits the number of interval fault patterns evaluated when constructing each transition probability. Instead of summing over every possible \(F_t\), the scalable method evaluates a selected set of fault patterns for the current candidate sequence, incoming hidden boundary state, and measured syndrome.
Once selected, each fault pattern is evaluated in the same way as in the exact method, using the same fault probability \(P(F_t\mid b_{t-1},h)\) and interval map \(\Psi_t^h\). The scalable approximation changes only which interval fault patterns are included in the sum.
Let
\[
\mathcal F_{t,h}^{\mathrm{select}}(b_{t-1},\mathbf y_t)
\]
denote the selected fault-pattern set, with
\[
\left| \mathcal F_{t,h}^{\mathrm{select}}(b_{t-1},\mathbf y_t) \right| \le B_F,
\]
containing at most \(B_F\) fault patterns for candidate sequence \(h\) in circuit interval \(t\).
The selected set may be generated using probability-informed or decoder-assisted search. Using the same fault probabilities and interval map as the exact method, the approximate joint syndrome-state transition probability is
\begin{equation}
\begin{aligned}
\widetilde K_t^h(\mathbf y_t,b_t\mid b_{t-1})
&=
\sum_{F_t\in \mathcal F_{t,h}^{\mathrm{select}}(b_{t-1},\mathbf y_t)}
P(F_t\mid b_{t-1},h)\\
&\quad\times\mathbf 1\!\left[
\Psi_t^h(b_{t-1},F_t)=(\mathbf y_t,b_t)
\right].
\end{aligned}
\label{eq:approx_transition_probability}
\end{equation}
We refer to \(\widetilde K_t^h\) as the approximate transition probability below. Figure~\ref{fig:scalable_inference}(a) illustrates the selection of a subset of the interval fault patterns \(F_t^1,F_t^2\) before their probabilities are summed.
Fault patterns outside the selected set are omitted from this sum. Because \(\widetilde K_t^h\) sums a subset of the nonnegative terms in the exact transition probability,
\[
0\le \widetilde K_t^h(\mathbf y_t,b_t\mid b_{t-1})
\le K_t^h(\mathbf y_t,b_t\mid b_{t-1}).
\]

Next, the scalable method limits the number of hidden boundary states propagated to the next circuit interval, as illustrated in Figure~\ref{fig:scalable_inference}(b). The scalable method first propagates the retained states, sums probabilities associated with the same outgoing hidden boundary state, and then retains at most \(B_S\) states for the next interval.
Let \(\mathcal B_t^h\) denote the retained boundary-state set for candidate sequence \(h\) at syndrome boundary \(t\). The retained set contains at most $B_S$ hidden boundary states:
\[
|\mathcal B_t^h|\le B_S.
\]
The forward probability before boundary-state selection and the approximate forward probability are
\begin{equation}
\begin{aligned}
\widetilde\alpha_t^{\mathrm{pre}}(b_t;h)
&=\sum_{b_{t-1}\in\mathcal B_{t-1}^h}
\widetilde\alpha_{t-1}(b_{t-1};h)
\widetilde K_t^h(\mathbf y_t,b_t\mid b_{t-1}),\\[3pt]
\widetilde\alpha_t(b_t;h)
&=\mathbf1\{b_t\in\mathcal B_t^h\}
\widetilde\alpha_t^{\mathrm{pre}}(b_t;h).
\end{aligned}
\label{eq:approx_forward_probabilities}
\end{equation}
The first line propagates the retained incoming states and merges probabilities associated with the same outgoing hidden boundary state. For probability-ranked boundary-state selection, the outgoing states are ranked using \(\widetilde\alpha_t^{\mathrm{pre}}(b_t;h)\) before the retained set is chosen. The second line retains the selected outgoing hidden boundary states for the next circuit interval.
Repeating it across the syndrome boundaries gives the approximate candidate likelihood
\[
\widetilde P(Y\mid h)=\sum_{b_T}\widetilde\alpha_T(b_T;h).
\]
The two budgets control different parts of the scalable method. \(B_F\) determines how many interval fault patterns are evaluated when constructing each approximate transition probability, while \(B_S\) determines how many hidden boundary states are carried between circuit intervals.

Finally, after computing the approximate candidate likelihood for each candidate sequence, the method combines candidates that share the same recovery target as shown in Figure~\ref{fig:scalable_inference}(c). The approximate unnormalized target probability is
\[
\widetilde Z_c(Y)=\sum_{h:\tau(h)=c}\pi(h)\widetilde P(Y\mid h).
\]
When
\[
\sum_{d\in\mathcal C}\widetilde Z_d(Y)>0,
\]
normalization gives the approximate target probability
\[
\widetilde P(c\mid Y)=\frac{\widetilde Z_c(Y)}{\sum_{d\in\mathcal C}\widetilde Z_d(Y)}.
\]
As in the exact method, probabilities from candidate sequences that share a recovery target are summed before the target is selected.

When valid lower and upper bounds on the exact unnormalized target probability are available, they can determine whether additional computation could change the selected recovery target:
\[
Z_c^-\le Z_c(Y)\le Z_c^+.
\]
Because the selected fault patterns and retained boundary states preserve nonnegative probability terms without reweighting, $\widetilde Z_c(Y)$ provides a lower bound on $Z_c(Y)$. A valid upper bound additionally requires accounting for omitted probability mass. A sufficient condition for certifying $c^\star$ as the exact MAP recovery target is
\[
Z_{c^\star}^->\max_{c\ne c^\star}Z_c^+.
\]
If the bounds do not determine the target, refinement can evaluate additional fault patterns, retain additional hidden boundary states, or replace an approximate interval calculation with the exact calculation when tractable. Calculations changed at an earlier syndrome boundary must then be propagated through the remaining circuit intervals. When valid bounds are unavailable, the method reports the approximate target probabilities directly.

For \(N_{\mathrm{cand}}=|\mathcal H|\) candidate sequences, \(T\) circuit intervals, at most \(B_S\) retained hidden boundary states per candidate and syndrome boundary, and at most \(B_F\) selected fault patterns per incoming state, a representative cost for evaluating one syndrome history is
\[
O\!\left(N_{\mathrm{cand}}\,T\,B_S\left[c_{\mathrm{gen}}+B_Fc_{\mathrm{eval}}\right]\right),
\]
where \(c_{\mathrm{gen}}\) is the cost of generating the selected fault-pattern set and \(c_{\mathrm{eval}}\) is the cost of evaluating one selected fault pattern.

Reuse can further reduce repeated computation. Candidate sequences that share the same compiled circuit up to a syndrome boundary can reuse the corresponding calculations when the required inputs are identical. Reuse itself does not discard any fault patterns or hidden boundary states, so the approximation remains controlled by fault-pattern selection and boundary-state selection.

\subsection{Logical Pauli Measurement Identification}
\label{subsec:method-pm}

The recovery target for logical Pauli measurement identification is the executed logical Pauli measurement \(M_P\), specified by the logical Pauli operator \(P\) and inferred from the classical measurement outcomes \(\mathbf o\). Unlike automorphism identification, which relies on the transport of data errors, gauging and ungauging can produce candidate-dependent correlations among auxiliary-qubit readouts and stabilizer measurements even without faults. We first describe these measurement correlations, then formulate the candidate likelihood under circuit faults, and finally present the scalable approximation.

\subsubsection{Ideal Measurement Correlations}
\label{subsubsec:ideal-pm-correlation}
For a logical Pauli measurement \(M_P\) described in Section~\ref{subsec:background-pm}, the classical measurement outcomes \(o\) include the syndromes measured before gauging and after ungauging, two rounds of Gauss, flux, and deformed stabilizer outcomes, and the auxiliary edge-\(Z\) readouts. The Gauss outcomes determine the logical measurement outcome \(\mu\), while the flux and deformed stabilizer outcomes measure checks of the gauged code. The deformed stabilizer checks and auxiliary edge readouts also constrain the syndrome measured after ungauging, providing a relation that depends on candidate \(M_P\). We first characterize these relations in the absence of faults.

A deformed stabilizer check has the form \(\widetilde S_j^{(P)} = S_j\prod_{e\in E_j^{(P)}}Z_e.\) When its measurement eigenvalue is \(+1\), measuring the auxiliary edges in the \(Z\) basis determines the eigenvalue of the corresponding original BB stabilizer \(S_j\). Let \(\omega_e\) be the binary outcome of edge \(e\), and let \(\omega\) collect these outcomes. With the prescribed check-sign convention in Section~\ref{subsec:background-pm}, the syndrome measured after ungauging, before applying the Pauli byproduct corrections, satisfies \((s_{\mathrm{post}})_j = \bigoplus_{e\in E_j^{(P)}}\omega_e.\) Collecting these relations gives 
\[
s_{\mathrm{post}}=Q_P\omega,
\]
where \(Q_P\) is a binary matrix whose \(j\)-th row specifies the auxiliary edges in \(E_j^{(P)}\). This provides the first relation between the candidate Pauli \(P\), the auxiliary edge outcomes, and the measured syndrome.

Other relations follow from the measured gauged-code checks. For a fault-free encoded input and the fixed sign convention, the two rounds of Gauss measurements agree, the flux and deformed stabilizer outcomes are zero, and the initial BB syndrome is zero. For the specified maximally mixed encoded input, the independent random bits underlying the Gauss and auxiliary edge outcomes are uniform. Let \(u\in\mathbb F_2^{d_P}\) collect these independent ideal measurement bits, where \(d_P\) is their number. Let \(r_P\) be a fixed binary offset and \(L_P\) the ideal outcome mapping matrix with \(d_P\) independent columns. The complete ideal classical measurement outcomes for candidate \(M_P\) can then be written as
\[
o_P^{\mathrm{ideal}}=r_P\oplus L_Pu.
\]
This affine law includes \(s_{\mathrm{post}}=Q_P\omega\) and the relations among the other classical measurement outcomes. Its \(2^{d_P}\) possible outcomes each have probability \(2^{-d_P}\). The random bits \(u\) represent intrinsic quantum measurement randomness rather than faults.

We define \(H_P\) as a binary matrix whose rows encode all ideal parity relations of candidate \(M_P\), including the relation between the auxiliary edge outcomes and the syndrome after ungauging, and the relations among the other measured outcomes. It satisfies
\[
H_PL_P=0,\qquad \ker H_P=\operatorname{im}L_P.
\]
Here \(\ker H_P\) is the null space of \(H_P\), and \(\operatorname{im}L_P\) is the column space of \(L_P\). The first equality ensures that all ideal classical measurement outcomes satisfy the parity relations, while the second ensures that these relations characterize all possible ideal classical measurement outcomes. Thus, the ideal candidate likelihood is
\[
P_{\mathrm{ideal}}(o\mid M_P) = 2^{-d_P}\mathsf{1}\!\left[ H_P(o\oplus r_P)=0 \right].
\]

\subsubsection{Exact Pauli Measurement Identification}
\label{subsubsec:pauli-exact}
We next compute the candidate likelihood $P(\mathbf o\mid M_P)$ under circuit faults. The exact method first determines the parity-check violation that must be produced by faults for candidate \(M_P\), then sums the probabilities of all complete fault patterns producing this violation.

For the fixed Clifford gauging and ungauging circuit with independent stochastic Pauli, reset, and measurement faults, let \(\delta_P(F)\) denote the fault-induced change in classical measurement outcomes under complete fault pattern \(F\). This change is independent of the ideal measurement bits \(u\), giving
\[
o=r_P\oplus L_Pu\oplus\delta_P(F).
\]

First, the method determines the parity-check violation for candidate \(M_P\). Since \(H_PL_P=0\), applying \(H_P\) eliminates the contribution of the independent ideal measurement bits \(u\):
\[
\begin{aligned}
z_P(o) &\equiv H_P(o\oplus r_P)\\ &=H_PL_Pu\oplus H_P\delta_P(F)\\ &=H_P\delta_P(F).
\end{aligned}
\]
For each candidate \(M_P\), the classical measurement outcomes \(o\) determine the parity-check violation \(z_P(o)\) that must be produced by the complete fault pattern.

Next, the method sums the probabilities of all complete fault patterns producing a specified parity-check violation \(z\). Define
\[
q_P(z)= \sum_F P(F\mid M_P) \mathsf{1}\!\left[ H_P\delta_P(F)=z \right].
\]
This probability combines the contributions of individual fault locations. For candidate \(M_P\), let \(j=1,\ldots,N\) index the independent modeled fault locations. At location \(j\), let \(\mathcal A_j\) contain the no-fault outcome and all allowed fault outcomes, with probability \(P_j(a)\) for \(a\in\mathcal A_j\). Let \(F_{j,a}\) denote the single-location fault pattern with outcome \(a\) at location \(j\). Its parity-check change is
\[
v_{P,j,a}=H_P\delta_P(F_{j,a}).
\]
For a complete fault pattern \(F\) with outcome \(a_j\) at each fault location, the resulting parity-check violation is the XOR of the individual parity-check changes
\[
H_P\delta_P(F) = \bigoplus_{j=1}^{N}v_{P,j,a_j}.
\]
Let \(q_{P,j}(z)\) denote the probability that the first \(j\) fault locations produce the combined parity-check violation \(z\). The method computes this probability using
\[
\begin{aligned}
q_{P,0}(z)&=\mathsf{1}[z=0],\\ q_{P,j}(z)&= \sum_{a\in\mathcal A_j} P_j(a)\, q_{P,j-1}(z\oplus v_{P,j,a}).
\end{aligned}
\]
For each possible outcome \(a\) at location \(j\), its probability is combined with the probability of the preceding parity-check violation \(z\oplus v_{P,j,a}\). After all \(N\) fault locations, \(q_{P,N}(z)=q_P(z)\).

Finally, the method computes the candidate likelihood. For each complete fault pattern satisfying \(H_P\delta_P(F)=z_P(o)\), the equality \(\ker H_P=\operatorname{im}L_P\) ensures that exactly one choice of \(u\) produces the classical measurement outcomes \(o\). Since \(u\) is uniform, this choice has probability \(2^{-d_P}\). The exact candidate likelihood is therefore
\[
P(o\mid M_P) = 2^{-d_P}q_P\!\left(z_P(o)\right).
\]
After computing the candidate likelihoods, candidates are combined by recovery target as described in Section~\ref{subsec:method-model}. Exact computation can become expensive because the number of intermediate parity-check violations with nonzero probability may grow exponentially, motivating the scalable method below.

\subsubsection{Scalable Pauli Measurement Identification}
\label{subsubsec:pauli-scalable}
The scalable method approximates the candidate likelihood \(P(o\mid M_P)\) by limiting the fault combinations and intermediate probability states considered when computing \(q_P(z)\). The parity-change budget \(B_{\mathrm{PC},i}\) limits the number of nonzero parity-check changes considered in circuit stage \(i\). The violation-state budget \(B_V\) limits the number of intermediate probability states retained between stages.

First, the method divides the fault locations of the gauging and ungauging circuit into consecutive stages. Fault-pattern selection retains fault patterns with at most \(B_{\mathrm{PC},i}\) nonzero parity-check changes in stage \(i\). A fault may produce zero parity-check change, so faults with \(v_{P,j,a}=0\) are still included in the probability sum without using the budget. Thus, the parity-change budget \(B_{\mathrm{PC},i}\) does not limit the total number of faults. The retained fault patterns keep their original probabilities without renormalization.

Next, the method limits the intermediate probability states retained across stages. Each state contains an accumulated parity-check violation \(x\) and the count information needed to enforce the parity-change budgets. The method propagates the retained states, sums probabilities associated with identical outgoing states, and then retains at most \(B_V\) states for the next stage. To guide this selection, the classical measurement outcomes \(o\) determine the required final violation \(z=z_P(o)\). After accumulating \(x\), the remaining stages must produce the required remaining violation \(r=z\oplus x\). States that cannot produce \(r\) within the remaining parity-change budgets are excluded from the retained calculation.

For the remaining states, the method uses upper bounds on the probability of producing \(r\) to guide violation-state selection. Let \(P_{\mathrm{rem},\ell}^{+}(r)\) denote an upper bound on the probability that the remaining stages produce the bits of \(r\) in subset \(\ell\), under the remaining stage and count conditions. Since producing the complete violation \(r\) requires satisfying every such subset,
\[
P_{\mathrm{rem}}^{+}(r) = \min_{\ell}P_{\mathrm{rem},\ell}^{+}(r)
\]
is also an upper bound on the probability of producing \(r\). These bounds, together with the accumulated probabilities, guide which intermediate states are retained.

Finally, the method sums the original probabilities of the retained fault combinations producing the required violation \(z\). Denote this probability by \(\widetilde q_P(z)\). Because the retained probabilities are not renormalized and omitted contributions are nonnegative,
\[
0\le\widetilde q_P(z)\le q_P(z).
\]
The approximate candidate likelihood is therefore
\[
\widetilde P(o\mid M_P) = 2^{-d_P}\widetilde q_P\!\left(z_P(o)\right).
\]
The approximate candidate likelihoods are then aggregated by recovery target as in Section~\ref{subsec:method-model}.

Valid lower and upper bounds on \(q_P(z)\) must account for fault patterns excluded by the parity-change budgets and intermediate probability states not retained under \(B_V\). Let \(q_P^-(z)\) and \(q_P^+(z)\) denote these bounds. Then,
\[
\begin{aligned}
q_P^-(z)&\le q_P(z)\le q_P^+(z),\\ P^\pm(o\mid M_P) &= 2^{-d_P}q_P^\pm\!\left(z_P(o)\right).
\end{aligned}
\]
The candidate likelihood bounds can be aggregated into bounds on the exact unnormalized target probabilities, allowing exact-model MAP certification as described in Section~\ref{subsubsec:auto-scalable-inference}. If the bounds do not determine the recovery target, the method can evaluate additional fault combinations or retain additional intermediate probability states.

\subsection{T-Injection Identification}
\label{subsec:method-t}
The recovery target for T-injection identification is the executed instruction \(T_P\), specified by the logical Pauli axis \(P\) of its joint measurement \(M_{Z_T\otimes P}\).We first describe how the instruction is distinguished from a standalone logical Pauli measurement and then extend the Pauli measurement likelihood methods to T injection.

\subsubsection{T-Injection Boundaries and Pauli-Axis Identification}
\label{subsubsec:t-injection-axis}
T injection uses a separately prepared logical magic state, the joint logical Pauli measurement \(M_{Z_T\otimes P}\), and an additional destructive \(X_T\) measurement, as described in Section~\ref{subsec:background-t}. The additional measurement channel distinguishes the T-injection instruction from a standalone logical Pauli measurement, while its Pauli axis \(P\) remains unknown.

In the absence of faults, the joint measurement outcome \(m\) and the \(X_T\) measurement outcome \(n\) satisfy \(P(m,n\mid T_P)=\frac14.\) Their probabilities are independent of \(P\), so these two outcomes alone cannot identify the Pauli axis.

Therefore, identifying \(P\) uses the logical Pauli measurement identification method in Section III-C. The joint measurement \(M_{Z_T\otimes P}\) follows the same gauging procedure, except that one Gauss's law operator is multiplied by the factory logical operator \(Z_T\), so that the product of all Gauss's law operators corresponds to the joint logical Pauli operator \(Z_T\otimes P\).
The classical measurement outcomes of this joint measurement can therefore be used to construct candidate-dependent ideal parity relations as in Section~\ref{subsec:method-pm}. Let \(H_{T_P}\) denote the matrix of ideal parity relations for the complete T-injection circuit, with fixed binary offset \(r_{T_P}\) and \(d_{T_P}\) independent ideal measurement bits.
For the specified encoded input, the ideal classical measurement outcomes can be written as
\[
\mathbf o_{T_P}^{\mathrm{ideal}} = r_{T_P}\oplus L_{T_P}u,
\]
where \(u\in\mathbb F_2^{d_{T_P}}\) contains independent uniform ideal measurement bits and \(L_{T_P}\) is the ideal outcome mapping matrix with independent columns. As in Section~\ref{subsec:method-pm},
\[
H_{T_P}L_{T_P}=0, \qquad \ker H_{T_P}=\operatorname{im}L_{T_P}.
\]
The ideal candidate likelihood is
\[
P_{\mathrm{ideal}}(o\mid T_P) = 2^{-d_{T_P}} \mathsf1\left[ H_{T_P}(o\oplus r_{T_P})=0 \right].
\]

\subsubsection{Exact T-Injection Identification}
\label{subsubsec:t-injection-exact}
We extend the exact Pauli measurement identification method in Section~\ref{subsubsec:pauli-exact} to identification of T-injection instruction \(T_P\). The initial magic state is assumed to be prepared without faults, while the circuit noise model additionally includes faults during resource coupling and the destructive \(X_T\) measurement.

For a complete fault pattern \(F\), let \(\delta_{T_P}(F)\) denote the fault-induced change in classical measurement outcomes. Under the fixed circuit and stochastic Pauli fault model,
\[
\mathbf o = r_{T_P}\oplus L_{T_P}u \oplus\delta_{T_P}(F).
\]
Using the ideal parity relations \(H_{T_P}\) defined in Section~\ref{subsec:method-pm}, the parity-check violation is
\[
z_{T_P}(o) = H_{T_P}(o\oplus r_{T_P}) = H_{T_P}\delta_{T_P}(F).
\]
The exact method then sums the probabilities of all complete fault patterns producing a specified parity-check violation \(z\).
\[
q_{T_P}(z) = \sum_F P(F\mid T_P) \mathsf{1}\!\left[ H_{T_P}\delta_{T_P}(F)=z \right].
\]
The same fault-location probability calculation from Section~\ref{subsubsec:pauli-exact} applies to the complete T-injection circuit, including its additional resource-coupling and measurement locations. The exact candidate likelihood is
\[
P(o\mid T_P) = 2^{-d_{T_P}} q_{T_P}\!\left(z_{T_P}(o)\right).
\]

\subsubsection{Scalable T-Injection Identification}
\label{subsubsec:t-injection-scalable}
We next extend the scalable Pauli measurement identification method in Section~\ref{subsubsec:pauli-scalable} to T-injection identification. The parity-change budget \(B_{\mathrm{PC},i}\) limits the number of nonzero parity-check changes considered in each circuit stage, while the violation-state budget \(B_V\) limits the intermediate probability states retained between stages. Both selections account for the additional resource-coupling and resource-measurement fault locations.

Applying the same fault-pattern selection and violation-state selection gives the retained probability \(\widetilde q_{T_P}(z)\). The approximate candidate likelihood is
\[
\widetilde P(o\mid T_P) = 2^{-d_{T_P}} \widetilde q_{T_P}\!\left(z_{T_P}(o)\right).
\]

Valid lower and upper bounds on the exact candidate likelihood are computed by accounting for omitted probabilities as in Section~\ref{subsubsec:pauli-scalable}. These bounds can be aggregated by recovery target to support exact-model MAP certification and refinement as described in Section~\ref{subsubsec:auto-scalable-inference}.

\subsection{Logical-Circuit Identification}
\label{subsec:method-circuit}
Logical-circuit identification extends the preceding instruction identification methods to a logical circuit \(\Gamma(h)\) constructed from automorphisms, logical Pauli measurements, and T injections. The candidate likelihood \(P(o\mid h)\) must account for how preceding instructions affect subsequent classical measurement outcomes.

The three instruction methods are combined by propagating the accumulated data error, encoded state, and measurement-dependent corrections throughout the logical circuit. Automorphisms transport data errors through their physical symplectic actions as described. Logical Pauli measurements update the encoded state according to the measurement outcome \(\mu\) and produce Pauli byproducts. T injections use outcomes \(m,n\) to determine the conditional operation and corrections, which can affect subsequent measurement bases through classical feedforward.

To compute the exact candidate likelihood, let \(N_{\mathrm{inst}}\) denote the number of instructions in \(\Gamma(h)\). Let \(o_j\) denote the classical measurement outcomes associated with instruction \(j\), including its SEC outcomes when present, and let \(F_j\) denote the fault pattern in its corresponding physical circuit. The quantities \(o_{<j}\) and \(F_{<j}\) denote the classical measurement outcomes and fault patterns preceding instruction \(j\). The candidate likelihood can then be computed as
\[
P(o\mid h) = \sum_{F_1,\ldots,F_{N_{\mathrm{inst}}}} \prod_{j=1}^{N_{\mathrm{inst}}} P(F_j,o_j\mid F_{<j},o_{<j},h).
\]

Each conditional probability is evaluated using the encoded state, accumulated data error, and corrections resulting from preceding instructions. When classical feedforward changes subsequent operations or measurement bases, the corresponding physical circuit and fault probabilities are determined by the preceding measurement outcomes. This preserves the dependencies between instructions while summing over complete fault patterns. These conditional probabilities cannot generally be replaced by independently computed single-instruction candidate likelihoods.

Scalable identification approximates the same candidate likelihood \(P(\mathbf o\mid h)\) by applying fault-pattern selection and boundary-state selection to automorphism-only logical circuits, and fault-pattern selection and violation-state selection across the continuous physical circuit containing logical Pauli measurements and T injections. The intermediate probabilities must preserve the information needed to evaluate subsequent instructions, including measurement-dependent corrections and changes to the encoded state. The resulting approximate candidate likelihoods \(\widetilde P(o\mid h)\) are combined by recovery target to obtain \(\widetilde Z_c(o)\), as described in Section~\ref{subsec:method-model}. Valid probability bounds and refinement can be applied as described in Sections~\ref{subsubsec:auto-scalable-inference} and~\ref{subsubsec:pauli-scalable}.

\section{Experimental Setup}
\label{sec:experimental-setup}
This section specifies the codes, recovery targets, candidate sets, measurement schedules, noise models, and inference settings used to evaluate logical-operation identification. We first specify the operation space and recovery targets, then describe how classical measurement outcomes are collected and how circuit faults are modeled. Finally, we summarize the inference configurations and evaluation criteria shared across experiments. The experiments were run on 8, 16, or 32 CPU cores with Intel Xeon Gold 6338 processors at 2.0 GHz, with up to 72 GiB of allocated memory per run.

\subsection{Codes and Recovery Targets}
We evaluate logical-operation identification on two bivariate bicycle codes, BB18 \([[18,4,2]]\) and BB72 \([[72,12,6]]\). The experiments include automorphism identification on BB18, as well as logical Pauli measurement, T-injection, and logical-circuit identification on both BB18 and BB72.

For automorphism identification, an executed automorphism $g$ induces a logical Clifford gate, represented by its logical action \(\Lambda_g\), and is realized by a compiled physical circuit. The exact automorphism experiments identify either the executed automorphism \(g\) or an ordered automorphism sequence \(h\), and separately evaluate their corresponding logical actions \(\Lambda_g\) and \(\Lambda_h\).
For logical Pauli measurements and T injections, the recovery targets are the executed instructions \(M_P\) and \(T_P\), respectively.
The scalable-inference evaluation of logical circuits uses the logical circuit \(\Gamma(h)\) as the primary recovery target. We additionally report the net logical action \(\Lambda_h\) for automorphism-based logical circuits when different logical circuits induce the same final action.

\subsection{Candidate Sets and Circuit Workloads}
We next define the candidate sets and logical circuits used in the experiments. The sets and circuits below serve only as test instances. Identification is performed independently for each executed instance rather than by selecting the best-matching candidate from the test set. 

Automorphism identification progresses from single automorphisms to ordered automorphism sequences. For single automorphism identification with exact inference on BB18, we use a five-automorphism set $\{I,x,x^{-1},y,y^{-1}\}$, which induces three distinct net logical actions. For ordered automorphism sequence identification with exact inference on BB18, we use the same five-automorphism set and evaluate all \(5^4=625\) ordered sequences of sequence length \(q=4\). We then evaluate a broader set of 216 automorphisms that induces 36 distinct logical actions. We sample 200 ordered automorphism sequences with \(q=4\) from this broader set.

For logical Pauli measurement identification, we evaluate all 15 nonidentity two-qubit Pauli measurements \(M_P\) on two fixed logical ports of BB18 and BB72. For T-injection identification, we evaluate six candidate instructions \(T_P\) on the same codes, with \(P\in\{XI,YI,ZI,IX,IY,IZ\}.\) Both instruction families use exact ideal candidate likelihoods in the absence of faults and scalable inference under circuit noise.

For logical-circuit identification with scalable inference, we first evaluate automorphism-based logical circuits that contain automorphisms and SEC rounds, without logical Pauli measurements or T injections. We test four workloads on both BB18 and BB72, including the eight-gate A/B circuit workload (AB), Greenberger-Horne-Zeilinger (GHZ), Bernstein-Vazirani (BV), and the restricted \(\mathbb Z_2^2\) coordinate-subgroup hidden subgroup problem (HSP). AB contains 16 eight-gate circuits, while GHZ, BV, and HSP each contain four two-gate circuits. We additionally evaluate Deutsch-Jozsa (DJ) on BB18 using four directly compiled encoded quadratic-phase oracle circuits rather than the automorphism candidate set.

We then extend the logical-circuit evaluation to a non-Clifford Pauli-rotation circuit that combines all three instruction types. The logical circuit follows \(\mathrm{AUTO}(g)\rightarrow T_P\rightarrow\mathrm{AUTO}(g^{-1})\rightarrow M_{Q}\), where the final logical Pauli measurement is adapted according to preceding measurement outcomes. The eight candidates combine two initial automorphism choices \(I\) and \(g\), two T-injection axes \(P\in\{ZI,YI\}\), and two initial logical Pauli measurement choices \(Q\in\{XI,IX\}\).

\subsection{Syndrome Extraction and Measurement Schedules}
Identification depends on which intermediate syndrome information is retained, so we evaluate schedules with different syndrome-extraction schedules. With syndrome extraction after every automorphism, an SEC is applied at entry and after every automorphism,
\[
\mathrm{SEC}_0 \rightarrow g_1 \rightarrow \mathrm{SEC}_1 \rightarrow g_2 \rightarrow \cdots \rightarrow g_q \rightarrow \mathrm{SEC}_q.
\]
In the endpoint-only schedule, syndrome extraction is performed only at entry and after the full sequence,
\[ \mathrm{SEC}_0 \rightarrow g_1 \rightarrow g_2 \rightarrow \cdots \rightarrow g_q \rightarrow \mathrm{SEC}_1 .
\]
The ordered automorphism sequence experiments additionally vary the number and placement of internal SEC boundaries to determine how syndrome-extraction cadence affects ordered-sequence resolution.

The logical Pauli measurement experiments use an initial BB SEC, two rounds of Gauss, flux, and deformed stabilizer measurements, auxiliary edge-\(Z\) readouts during ungauging, and a final BB SEC. The BB syndromes after unguaging are recorded before candidate-dependent Pauli byproduct corrections. Both ideal and noisy identification use the initial and final BB stabilizer measurement outcomes together with the auxiliary edge-\(Z\) readouts in Section~\ref{subsec:eval-pm}. We additionally compares identification using BB syndromes, intermediate deformed syndromes, and complete auxiliary measurement outcomes in Section~\ref{subsubsec:pm_identifiability}.

T-injection experiments extend the corresponding joint logical Pauli measurement with a final destructive factory-\(X_T\) measurement. In the ideal and circuit-level noise tests, we compare identification using the BB stabilizer and auxiliary edge-\(Z\) outcomes with identification using all classical outcomes, including the Gauss, flux, deformed stabilizer, and factory-\(X_T\) measurements.

The automorphism-based logical-circuit experiments use a fixed entry SEC followed by an SEC after each logical gate. The non-Clifford Pauli-rotation circuit uses a continuous physical execution with SEC and the corresponding logical measurement outcomes.

\subsection{Noise Models and Diagnostic Conditions}
We evaluate identification under several stochastic noise models and two diagnostic conditions. Automorphism identification relies on the transport of data errors, so we additionally evaluate how different error channels and error rates affect identification. The other instruction identifications are evaluated using a common circuit-level noise model.

\subsubsection{Noise models for exact automorphism inference}
The single-automorphism experiment uses an ideal one-interval setting with ideal state preparation and syndrome extraction. We compare a zero-error input with a known \(Y\) error on data qubit 0, denoted \(Y_0\), to determine whether the five automorphisms produce distinguishable syndromes.
The ordered automorphism-sequence experiments then introduce stochastic noise using simplified data and measurement models. State preparation and the gate operations in the SEC are ideal. Measurement outcomes are flipped only in the settings that explicitly include measurement errors. We consider the following five noise models.
\begin{enumerate}
    \item \textbf{Independent data bit-flip and phase-flip noise.}
    Independent $X$ and $Z$ errors are applied to the data qubits.

    \item \textbf{Single-qubit depolarizing data noise.}
    Data errors are drawn from the nonidentity single-qubit Pauli operators.

    \item \textbf{Measurement-error-only noise.}
    Recorded syndrome outcomes are flipped without introducing data errors.
    \item \textbf{Phenomenological data and measurement noise.}
    Independent data errors and measurement-outcome flips are both present.

    \item \textbf{Pauli faults in compiled automorphism circuits.}
    Pauli faults occur at registered CNOT and idle locations inside the compiled automorphism sequences. State preparation and syndrome extraction remain ideal.
\end{enumerate}
For the five-automorphism \(q=4\) experiment, we evaluate twelve nominal error rates from \(10^{-5}\) to \(2\times10^{-2}\). For the 216-automorphism experiment, we evaluate \(10^{-5}\), \(10^{-4}\), \(10^{-3}\), \(10^{-2}\), and \(2\times10^{-2}\). Each condition uses 1,400 shots. The nominal parameter \(p\) has model-specific meaning, so equal values of \(p\) do not imply equal total fault probability across the five models.

\subsubsection{Circuit noise model for other inferences}
The logical Pauli measurement, T-injection, and logical-circuit experiments use the same independent, gate-dependent stochastic Pauli model. Fault locations include one-qubit gates, CNOT gates, resets, measurements, and modeled idles. One-qubit gates and idles use uniformly distributed nonidentity single-qubit Paulis. CNOT faults use the fifteen nonidentity two-qubit Paulis. Reset faults flip the prepared state, and measurement faults flip the recorded outcome. The logical \(T\) state is prepared without faults. We evaluate three error-rate profiles, LOW, NOMINAL, and HIGH, as summarized in Table~\ref{tab:circuit_noise}. Each evaluated noise cohort contains 1024 shots. We additionally evaluate the ideal logical Pauli measurement and T-injection experiments in the absence of faults.

\begin{table}[t]
\centering
\caption{Full-circuit fault probabilities per location. Idle rates
apply per modeled idle unit.}
\label{tab:circuit_noise}
\footnotesize
\setlength{\tabcolsep}{4.5pt}
\begin{tabular}{lccc}
\hline
Location & LOW & NOMINAL & HIGH \\
\hline
One-qubit gate
    & $10^{-5}$
    & $10^{-4}$
    & $3\times10^{-4}$ \\
CNOT
    & $10^{-4}$
    & $10^{-3}$
    & $3\times10^{-3}$ \\
Reset
    & $5\times10^{-5}$
    & $5\times10^{-4}$
    & $1.5\times10^{-3}$ \\
Measurement
    & $10^{-4}$
    & $10^{-3}$
    & $3\times10^{-3}$ \\
Idle
    & $10^{-5}$
    & $10^{-4}$
    & $3\times10^{-4}$ \\
\hline
\end{tabular}
\end{table}

\subsubsection{PASSIVE and MARKER conditions}
For the automorphism, automorphism-based logical-circuit, and DJ experiments, we evaluate two diagnostic conditions, PASSIVE and MARKER. The non-Clifford Pauli-rotation circuit is evaluated under MARKER.

PASSIVE introduces no deliberate data error. Each shot is generated only under the declared stochastic noise model. PASSIVE therefore includes fault-free shots as well as shots in which naturally occurring errors may provide information about the executed operation.
MARKER additionally inserts a known Pauli data error before the logical operations while preserving the same stochastic noise model. The marker provides a controlled input whose transport is reflected in subsequent measured syndromes.
PASSIVE and MARKER therefore use the same underlying stochastic noise model and differ only in whether a known Pauli data error is deliberately introduced.
% This comparison isolates the effect of informative error transport from limitations caused by syndrome information and the inference procedure itself.

\subsection{Inference Settings}
The exact automorphism evaluation first compares Exact-Sum with Exact-Max, as defined in Section~\ref{subsubsec:auto-exact-inference}, and then uses Exact-Sum for the remaining exact automorphism experiments.

For logical Pauli measurement and T-injection identification, the ideal experiments use exact ideal candidate likelihoods, while the noisy experiments use scalable inference with the parity-change budget \(B_{\mathrm{PC},i}\) and violation-state budget \(B_V=4096\), as defined in Section~\ref{subsec:method-pm} and~\ref{subsec:method-t}. The parity-change budgets are fixed for each compiled circuit and noise profile. For BB72 logical Pauli measurements under NOMINAL and HIGH noise, we additionally evaluate refined violation-state selection using the same shots with unchanged parity-change budgets and violation-state budget.

The scalable automorphism-based logical-circuit inference evaluation uses a common baseline configuration across codes, workloads, and noise profiles. Within each block consisting of preparation or one logical gate together with the following SEC, the baseline considers at most two faulty physical locations, evaluates at most 256 selected fault patterns, and propagates at most 1,024 hidden boundary states. For BB72 under the HIGH noise profile, we additionally evaluate broader scalable configurations. The broader configuration allows more than two faulty locations and evaluates up to 8192 fault patterns, while continuing to propagate at most 1024 hidden boundary states. ``Through SEC0'', ``Through SEC1'', ``Through SEC2'', and ``Through SEC8'' cumulatively apply the broader multi-fault calculation through the indicated SEC boundary, while subsequent circuit blocks retain the baseline configuration. The physical circuits, noise model, syndrome histories, candidate set, priors, and recovery target remain unchanged across these comparisons.

The non-Clifford Pauli-rotation circuit uses the joint candidate likelihood defined in Section~\ref{subsec:method-circuit} with \(B_V=4096\) and the fixed parity-change budgets.

\subsection{Evaluation Metrics}
We evaluate identification accuracy for each recovery target and further examine how fault timing affects the availability of distinguishing information under PASSIVE.

Identification accuracy is the percentage of correct recovery-target identifications. For ideal identification, Bayes accuracy is the expected correctness of selecting a recovery target with the highest target probability. Strict unique identification counts an identification as correct only when the executed recovery target is the only recovery target with the highest target probability. For scalable inference, identification accuracy is based on approximate candidate likelihoods and does not represent exact Bayes accuracy under circuit noise. Identification is unavailable when all approximate candidate likelihoods are zero.

Beyond overall identification accuracy, we examine when distinguishing information becomes available during PASSIVE circuit execution.
PASSIVE retains all shots generated by the declared stochastic noise model, including fault-free shots and shots in which the first fault occurs only after part of the logical circuit has already executed. A late fault can provide identifying syndrome information for subsequent logical gates but cannot reveal gates that were completed before that fault occurred.
We therefore perform a retrospective fault-timing analysis for the automorphism-based experiments to determine which part of the circuit remains identifiable after a naturally occurring fault. \textbf{FIRST\_ANY} denotes the first fault anywhere in the compiled circuit. \textbf{FIRST\_LOGICAL} denotes the first fault within a logical operation block, including modeled idles belonging to that block. For either alignment, we exclude the logical gate containing the fault and evaluate only logical gates executed strictly afterward. Likelihoods are aggregated over candidates that share the same ordered logical gates strictly after the fault before the MAP decision. Shots with no logical gate remaining after the fault-containing gate are reported as not estimable.

\section{Evaluation}
\label{sec:evaluation}

\begin{figure}[t]
    \centering
    \includegraphics[width=\linewidth]{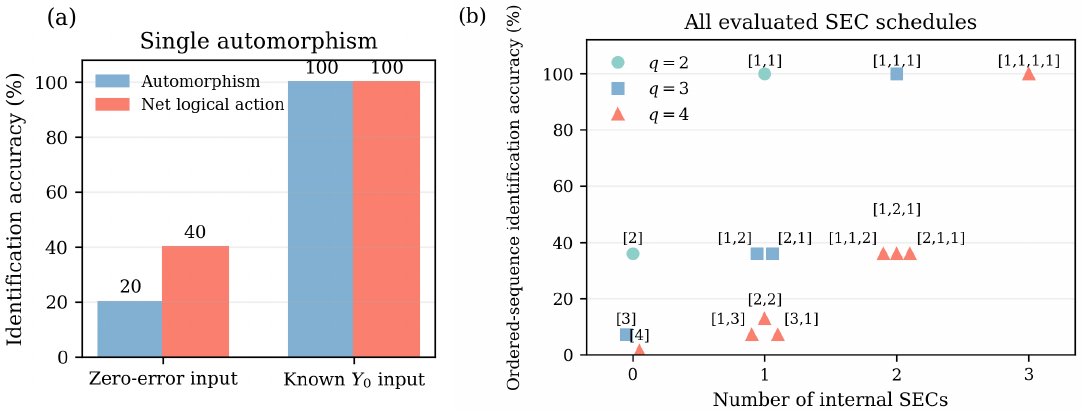}
    \caption{Exact identifiability on BB18. (a) Single-automorphism identification for the five-automorphism set \(\{I,x,x^{-1},y,y^{-1}\}\) under a zero-error input and a known \(Y_0\) error on data qubit 0, with uniform prior. (b) Ordered automorphism sequence identification for all \(5^q\) sequences with \(q=2,3,4\), using the same known \(Y_0\) input and varying the number and placement of internal SECs. Each labeled point corresponds to one evaluated SEC segmentation, where \([a,b,\ldots]\) denotes the numbers of automorphisms between successive SECs.}
    \label{fig:eval_auto_exact}
\end{figure}

\begin{figure*}[t]
    \centering
    \includegraphics[width=\textwidth]{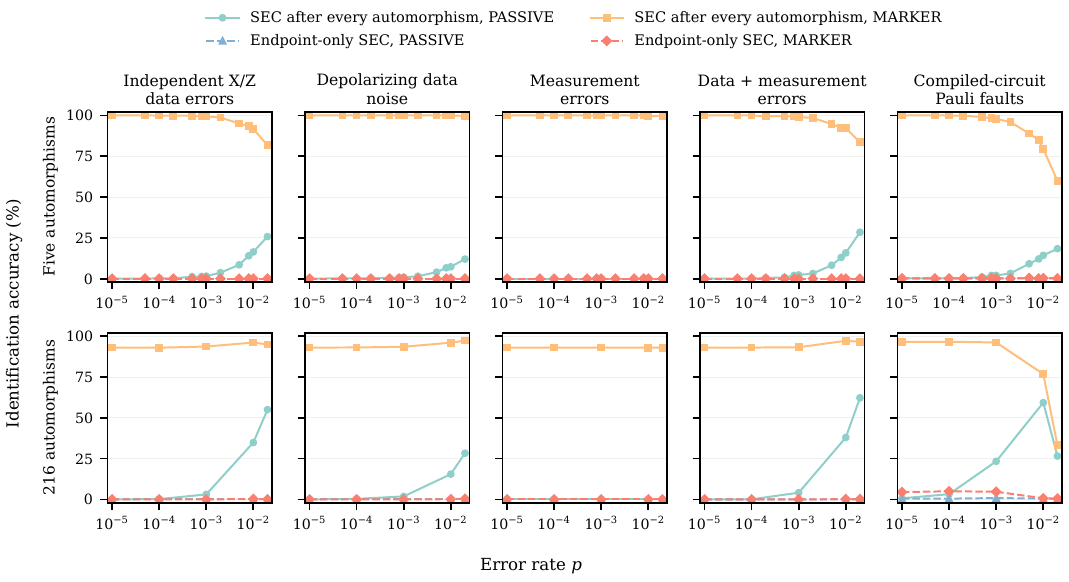}
    \caption{Ordered automorphism-sequence identification on BB18 with Exact-Sum under different noise models, syndrome-extraction schedules, and PASSIVE and MARKER conditions defined in Section~\ref{sec:experimental-setup}. The top row shows all \(q=4\) sequences formed from the five automorphisms. The bottom row shows the 200 sampled \(q=4\) sequences from the broader set of 216 automorphisms.}
    \label{fig:eval_auto_noisy1}
\end{figure*}

\begin{figure*}[t]
    \centering
    \includegraphics[width=\textwidth]{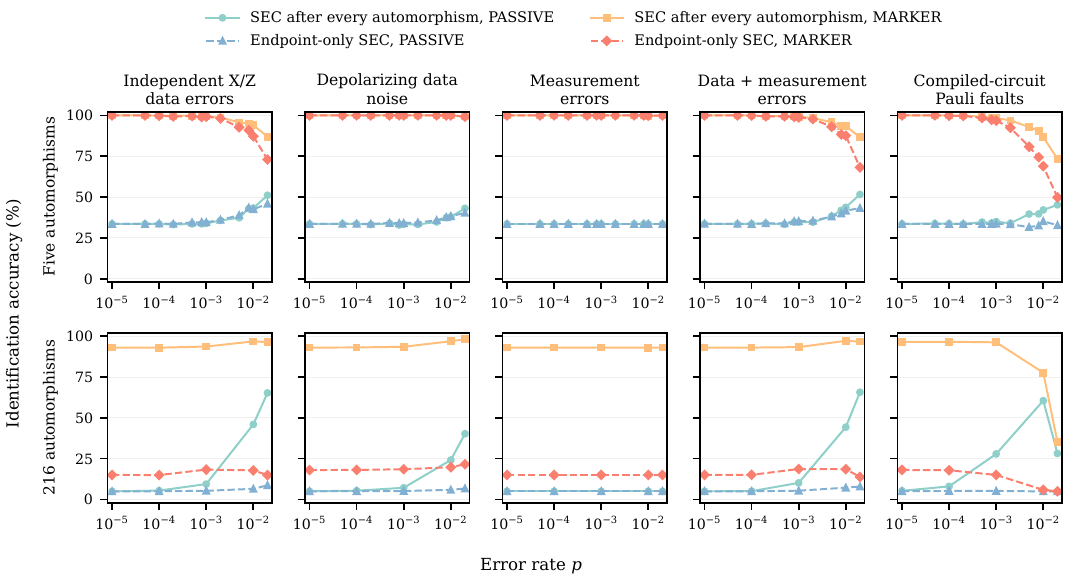}
    \caption{Net logical-action identification on BB18 with Exact-Sum for the same experiments as Figure~\ref{fig:eval_auto_noisy1}. The top row shows the five-automorphism experiment and the bottom row shows the 216-automorphism experiment. Prior-weighted candidate likelihoods from ordered sequences with the same net logical action are summed before the final decision.}
    \label{fig:eval_auto_noisy2}
\end{figure*}

\begin{figure*}[t]
    \centering
    \includegraphics[width=\textwidth]{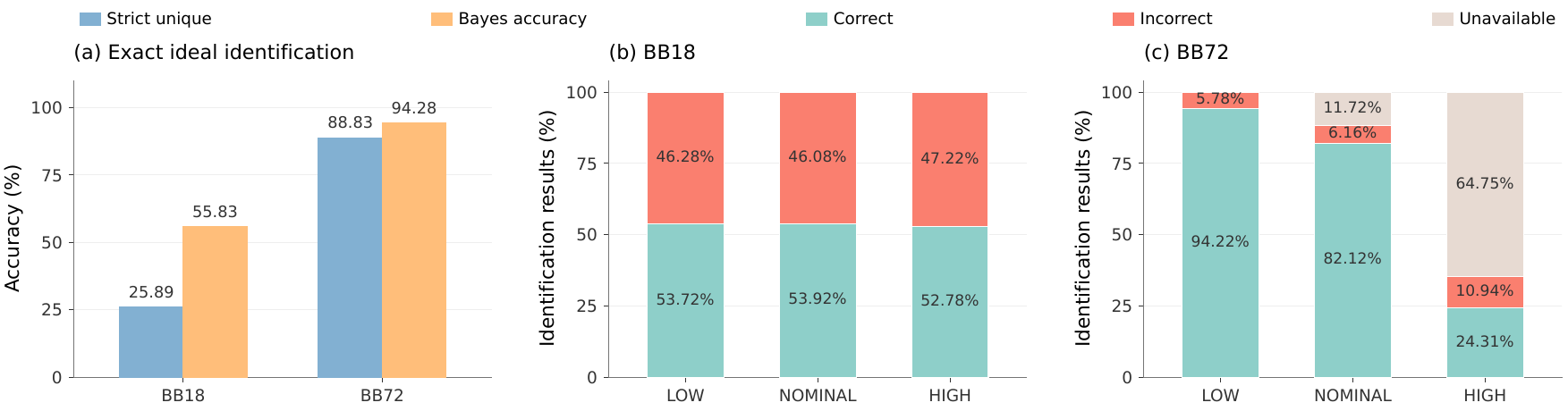}
    \caption{Logical Pauli measurement identification on BB18 and BB72.
    (a) Exact ideal identification among 15 logical Pauli measurements on two fixed logical ports. (b) and (c) Scalable identification under LOW, NOMINAL, and HIGH circuit-level noise on BB18 and BB72, respectively. The identification uses the initial and final BB stabilizer measurement outcomes together with the auxiliary edge-\(Z\) readouts as input. Stacked bars show the percentages of correct, incorrect, and unavailable identification, with each candidate weighted equally. Identification is unavailable when all candidate retained point likelihoods are zero.}
    \label{fig:eval_pm}
\end{figure*}

\begin{figure*}[t]
    \centering
    \includegraphics[width=\textwidth]{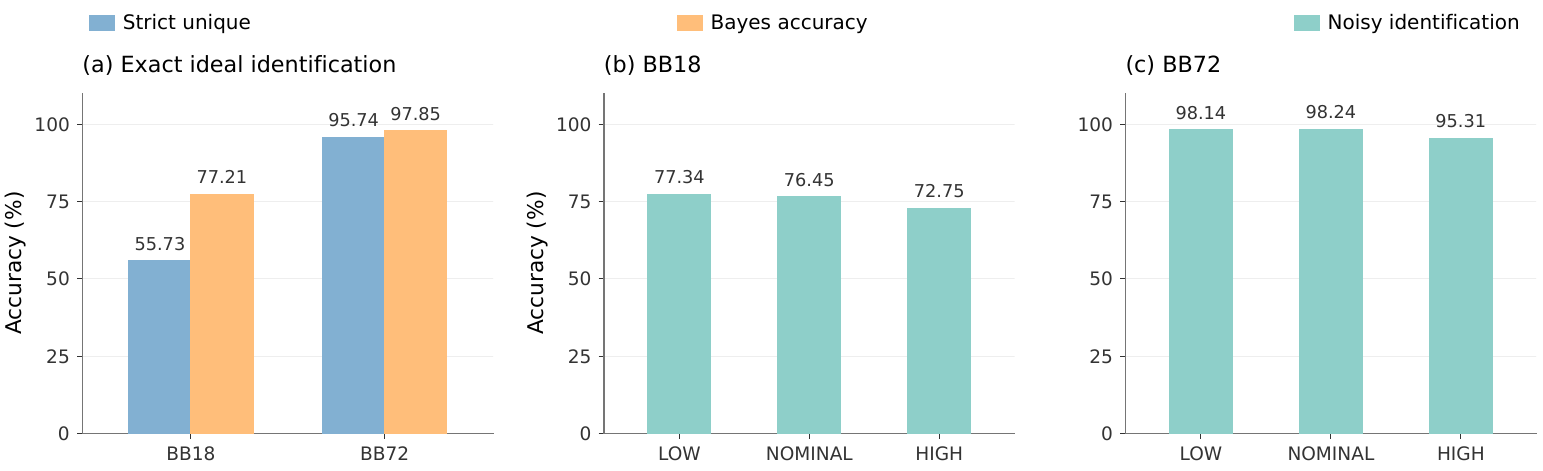}
    \caption{T-injection identification on BB18 and BB72.
    (a) Exact ideal identification among six T-injection instructions. (b) and (c) Scalable identification under LOW, NOMINAL, and HIGH circuit-level noise on BB18 and BB72, respectively. The identification uses the initial and final BB stabilizer measurement outcomes together with the auxiliary edge-\(Z\) readouts as input.}
    \label{fig:eval_t}
\end{figure*}

\begin{figure*}[t]
\centering
\includegraphics[width=\textwidth]{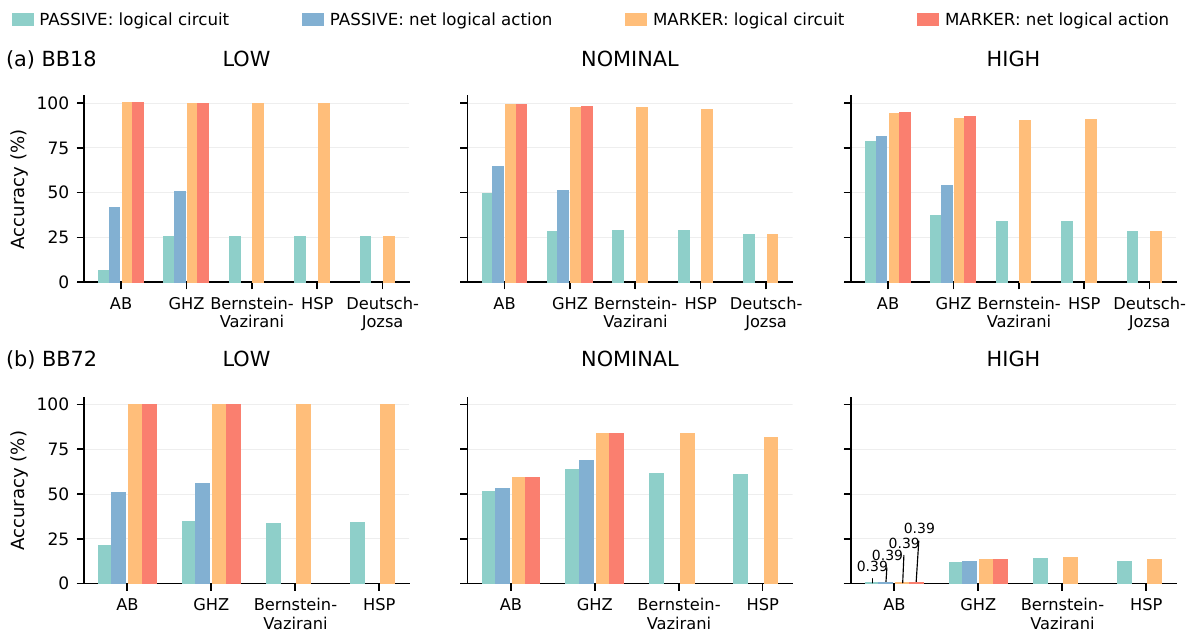}
\caption{Automorphism-based logical-circuit identification on BB18 and BB72 using the baseline scalable inference configuration defined in Section~\ref{sec:experimental-setup}. (a) BB18 identification results for AB, GHZ, BV, HSP, and the separately compiled DJ workload. (b) BB72 identification results for AB, GHZ, BV, and HSP. Columns show the LOW, NOMINAL, and HIGH noise profiles defined in Section~\ref{sec:experimental-setup}. Bars report logical-circuit identification under PASSIVE and MARKER, together with net logical-action identification for AB and GHZ. For the remaining workloads, the two recovery targets coincide. Refinement of BB72 HIGH results is discussed in Section~\ref{subsec:scalable_approximation}.}
\label{fig:eval_circuit_auto}
\end{figure*}

\begin{figure}[t]
    \centering
    \includegraphics[width=\columnwidth]{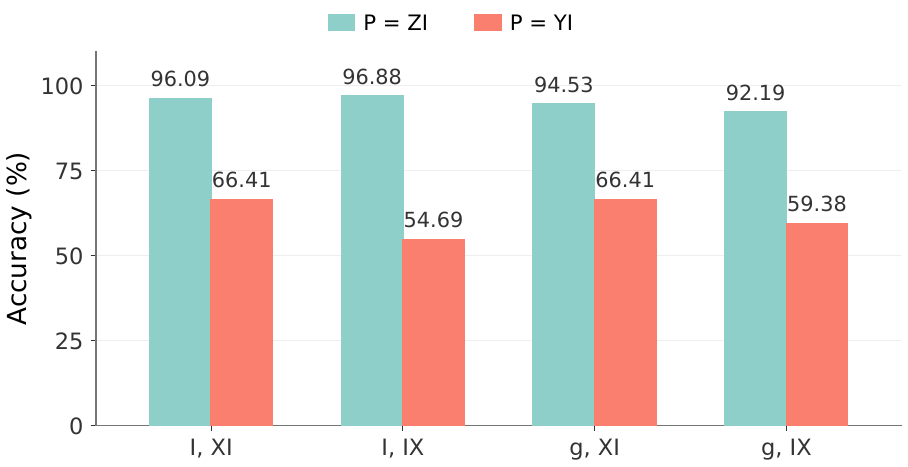}
    \caption{Logical-circuit identification combining all three instruction types on BB18. The tested circuit is a non-Clifford Pauli-rotation circuit as defined in Section~\ref{sec:experimental-setup}. Bars show logical-circuit identification accuracy for eight candidates under LOW circuit-level noise with MARKER. The horizontal axis specifies the candidate automorphism (\(I\) or \(g\)) and candidate logical Pauli measurement choice (\(Q=XI\) or \(IX\)), while colors distinguish the T-injection candidates (\(P=ZI\) or \(YI\)).}
    \label{fig:eval_circuit_mix}
\end{figure}

We evaluate logical-operation identification across automorphisms, logical Pauli measurements, T injections, and logical circuits on BB18 and BB72. We first establish the information required for exact automorphism identification and evaluate its robustness across noise models and syndrome-extraction schedules. We then examine ideal and noisy identification of logical Pauli measurements and T injections. Finally, we evaluate scalable identification of logical circuits on both codes.

\subsection{Automorphism Identification}
\label{subsec:eval-auto}

\subsubsection{Exact Identification}
We begin with exact inference on BB18 to determine what information can be recovered from a syndrome history before varying the noise model. The experiments establish three principles for automorphism identification.
% \Shifan{begin enumerate here}
\begin{enumerate}
    \item \textbf{Distinguishable information from data errors.} Identification requires a data error whose syndromes depend on the executed automorphisms. 
    \item \textbf{Intermediate syndrome extraction.} Intermediate SECs are needed to distinguish ordered automorphism sequences that produce the same endpoint syndrome.
    \item \textbf{Fault marginalization.} Candidate likelihoods account for multiple fault explanations of the same syndrome history by summing probabilities over all compatible fault patterns rather than retaining only the most likely one.
\end{enumerate}
% Identification requires a data error whose syndromes depend on the executed automorphisms. Intermediate syndrome extraction is needed when the ordered automorphism sequence is the recovery target. Finally, identification should marginalize over compatible complete fault patterns rather than retain only the most likely one.

We first ask when a single automorphism can be distinguished from its syndrome. Figure~\ref{fig:eval_auto_exact}(a) compares the five-automorphism set under an ideal one-interval experiment as defined in Section~\ref{sec:experimental-setup}. With a zero-error input, all five automorphisms produce the same syndrome and therefore cannot be distinguished. If the five automorphisms are equally likely, uniform guessing gives 20\% automorphism-identification accuracy. The five automorphisms induce three net logical actions with multiplicities one, two, and two, so always choosing one of the two most common logical actions gives 40\% net logical-action accuracy.
A known \(Y_0\) error instead produces five distinct syndromes. The automorphism and its net logical action can therefore both be identified from the syndrome. This comparison shows that an error provides identifying information only when different automorphisms transform it into distinguishable syndromes.

We next ask whether the syndrome history can identify the ordered automorphism sequence rather than only its net logical action. Figure~\ref{fig:eval_auto_exact}(b) illustrates identification accuracy for the full set of \(5^q\) ordered automorphism sequences with \(q=2,3,4\) under different syndrome-extraction schedules with a known $Y_0$ error. With endpoint-only syndrome extraction, many distinct sequences produce the same final syndrome. The resulting ordered automorphism sequence accuracy is 36\% for \(q=2\), 7.2\% for \(q=3\), and 1.44\% for \(q=4\).

For \(q=4\), adding one internal SEC increases the accuracy to 7.2\%--12.96\%, depending on its placement. Adding two internal SECs increases it to 36\%. Syndrome extraction after every automorphism distinguishes all 625 sequences and gives 100\% ideal accuracy. These results show that intermediate syndrome measurements provide information needed to identify the ordered automorphism sequence when endpoint-only observation is insufficient.

Finally, we compare Exact-Max and Exact-Sum, as defined in Section~\ref{subsubsec:auto-exact-inference}, on selected cases from the five-automorphism experiments. A shot is selected when approximate inference and Exact-Sum have no highest-scoring net logical action in common. The following percentages are calculated over these selected shots.
% \Shifan{cannot understand here. What is the definition of diagnostic subset, ididn't see it in experimental setup sec, the following sentence is also confusing}
Exact-Max uniquely identifies the correct net logical action in 7.77\% of the selected shots, while giving the three logical actions the same scores in the remaining 92.23\% of shots.
Exact-Sum produces a unique decision for every shot and identifies the correct net logical action in 61.82\%. For each logical action, Exact-Max uses the most probable compatible complete fault pattern across the candidate automorphisms that induce that action. Exact-Sum instead adds the probability contributions from all compatible complete fault patterns and all candidate automorphisms that induce the same logical action. The large difference shows that fault marginalization is important when multiple complete fault patterns contribute appreciable probability. We therefore use Exact-Sum for the remaining exact-inference experiments.

\subsubsection{Exact Identification Robustness Across Noise Models}\label{subsec:eval-robustness}
We next evaluate Exact-Sum using the five-automorphism set and the broader 216-automorphism set across the stochastic noise models and SEC schedules defined in Section~\ref{sec:experimental-setup}. Figure~\ref{fig:eval_auto_noisy1} reports identification accuracy for the ordered automorphism sequence, while Figure~\ref{fig:eval_auto_noisy2} reports identification accuracy for the corresponding net logical action. Comparing the two figures shows how identification accuracy changes with both the noise model and the requested recovery target.

Both candidate sets achieve high identification accuracy across all five noise models when MARKER provides a distinguishing input and syndrome extraction is performed after every automorphism. For \(10^{-5}\le p\le10^{-3}\), the five-automorphism set achieves 97.86\% -100\% ordered-sequence accuracy and 98.36\%-100\%  net logical-action accuracy across all five noise models.  For the broader 216-automorphism set, both targets achieve 92.93\%-96.57\% accuracy over the same noise range. Thus, when a known error supplies distinguishing transport and the intermediate syndrome history is available, exact identification remains accurate across different error mechanisms.

Figure~\ref{fig:eval_auto_noisy1} and Figure~\ref{fig:eval_auto_noisy2} show that both candidate sets exhibit contrasting trends under PASSIVE and MARKER as the error rate increases. Under MARKER, the known $Y_0$ error already provides distinguishing information at low error rates. Additional faults during circuit execution can disturb the subsequent syndromes, so the MARKER curves with syndrome extraction after every automorphism begin near 100\% and generally decrease as $p$ increases. Under PASSIVE, no known error is supplied. At small $p$, many shots contain little or no transported data error information, so ordered-sequence identification begins near 0\%. As data errors become more common, some are transported differently by different automorphisms and provide identifying information, causing PASSIVE accuracy to increase. For example, under independent \(X/Z\) data errors, MARKER ordered-sequence accuracy for five-automorphism set decreases from 100\% at \(p=10^{-5}\) to 81.71\% at \(p=2\times10^{-2}\) while the corresponding PASSIVE accuracy increases from 0\% to 25.79\%. The broader set follows the same general trends, but its MARKER accuracy begins around 93\%–96\% rather than near 100\%. To understand this gap, we separately examined the ideal syndrome histories of the 200 sampled sequences and found only 193 distinct MARKER syndrome histories under syndrome extraction after every automorphism. Some sampled sequences therefore share identical ideal MARKER syndrome histories, which helps explain the lower initial MARKER accuracy. For several data-error models, MARKER accuracy then increases slightly around \(p=10^{-2}\), consistent with additional transported data errors supplying information that separates some of these otherwise ambiguous sequences.

Identification accuracy also strongly depends on the noise model for both candidate sets. Independent data bit-flip and phase-flip noise, single-qubit depolarizing data noise, phenomenological data and measurement noise, and Pauli faults in compiled automorphism circuits can all place errors on the data that later automorphisms transport. In contrast, measurement errors change measured syndrome bits but do not place a Pauli error on the data for a later automorphism to transport. PASSIVE ordered-sequence accuracy therefore remains 0\% throughout the measurement-error-only sweep, and PASSIVE net logical-action accuracy remains at 33.5\% for five-automorphism set and 5\% for the broader set as \(p\) changes. Under MARKER, the known data error continues to provide the transported information, so identification remains close to 100\% and 93\%, respectively, as the measurement-error rate changes. The main difference appears at higher rates of data noise, where the broader set exhibits stronger degradation. Under Pauli faults in compiled automorphism circuits, PASSIVE ordered-sequence accuracy for the five-automorphism set continues to increase from 14.29\% to 18.36\% as \(p\) increases from \(10^{-2}\) to \(2\times10^{-2}\), whereas the broader set decreases from 59.29\% to 26.57\% over the same range. This difference is consistent with the larger number of fault locations in the sampled sequences. In this compiled model, automorphisms are implemented through physical-qubit permutations using SWAP operations, with each SWAP decomposed into three CX gates. The five-automorphism sequences contain approximately 230 registered CX and idle fault locations per sequence on average, compared with 585 to 1197 for the sampled broader sequences. The greater fault exposure provides more opportunities for informative transported errors at moderate rates, but also for multiple faults to obscure these patterns at higher rates.

Finally, the recovery target can lead to substantially different identification accuracy under the same noise and syndrome-extraction conditions for both sets. Different ordered automorphism sequences can induce the same net logical action, allowing the net action to remain identifiable even when the executed sequence is unresolved. At low error rates under PASSIVE, ordered-sequence identification begins near 0\% for both sets, while net logical-action identification begins around 33\% for the five-automorphism set and 5\% for the broader set. The different net-logical accuracy arises because the five-automorphism sequences map to only three net logical actions, whereas the broader set induces 36 distinct logical actions.

The recovery-target difference is even clearer with endpoint-only syndrome extraction. At \(p=10^{-3}\) under MARKER, ordered-sequence identification is 0\% for both sets under independent data bit-flip and phase-flip noise, single-qubit depolarizing data noise, measurement-error-only noise, and phenomenological data and measurement noise. In contrast, net logical-action identification ranges from 98.79\% to 100\% for the five-automorphism set and 15.00\% to 18.64\% for the broader set. With the same endpoint syndrome information, the net logical action is therefore identifiable even though the ordered automorphism sequence is not. Endpoint-only syndrome extraction can preserve information about the net logical action while losing the intermediate information needed to determine operation order.

\subsection{Logical Pauli Measurement Identification}
\label{subsec:eval-pm}
We next evaluate logical Pauli measurement identification on BB18 and BB72. Unlike automorphism identification, which relies on the transport of data errors, gauging-based logical Pauli measurements can produce candidate-dependent correlations among classical measurement outcomes even without faults. We first evaluate exact identification in the absence of faults and then examine scalable identification under circuit-level noise.

\subsubsection{Ideal Identification}
Figure~\ref{fig:eval_pm}(a) shows the exact ideal identification accuracy using the method in Section~\ref{subsec:method-pm}. The strict unique identification accuracies are 25.89\% on BB18 and 88.83\% on BB72, while Bayes accuracy reaches 55.83\% and 94.28\%, respectively. For both codes, Bayes accuracy is higher because the same classical measurement outcomes can be consistent with multiple candidate measurements. Bayesian identification can therefore correctly select the executed Pauli measurement even when the available outcomes do not uniquely determine it.

Although all 15 candidate measurements produce distinct ideal measurement laws on both codes, this does not guarantee that every execution provides sufficient information to distinguish them. The substantially lower identification accuracy on BB18 than on BB72 reflects this limitation. The possible measurement outcomes of different candidates overlap substantially on BB18, while the smaller overlap on BB72 leads to higher identification accuracy and a much smaller difference between Bayes and strict unique identification. We examine these differences in more detail in Section~\ref{subsubsec:pm_identifiability}.

\subsubsection{Identification Under Noise}
Figures~\ref{fig:eval_pm}(b) and (c) show the identification results for the two codes under circuit-level noise. Similar to automorphism identification under MARKER, logical Pauli measurements provide distinguishing information even without stochastic faults, so additional faults can obscure information that is already present. However, the effect on identification accuracy differs substantially between the two codes. On BB18, accuracy remains nearly unchanged at 53.72\%, 53.92\%, and 52.78\% under LOW, NOMINAL, and HIGH noise, respectively. In contrast, BB72 achieves 94.22\% accuracy at LOW noise, but decreases to 82.12\% at NOMINAL and 24.31\% at HIGH. Thus, despite its higher ideal identification accuracy, BB72 exhibits much stronger degradation in baseline scalable identification as the noise rate increases.

The results also show that the degradation on BB72 is not just incorrect identification. As shown in Figures~\ref{fig:eval_pm}(b) and (c), identification is available for all BB18 shots across the three noise profiles. On BB72, the fraction of unavailable identifications increases from 0\% at LOW to 11.72\% at NOMINAL and 64.75\% at HIGH, while incorrect identifications account for 5.78\%, 6.16\%, and 10.94\%, respectively. Much of the reduction in BB72 accuracy therefore occurs when scalable inference fails to retain a positive candidate likelihood, rather than selecting an incorrect measurement. Section~\ref{subsec:scalable_approximation} examines this limitation through refinement on the same shots.

\subsection{T-Injection Identification}
\label{subsec:eval-t}
We then evaluate ideal and scalable identification of the Pauli axis \(P\) of the T-injection instruction \(T_P\) on BB18 and BB72, as illustrated in Figure~\ref{fig:eval_t}.

\subsubsection{Ideal Identification}
Figure~\ref{fig:eval_t}(a) presents the exact ideal identification accuracy for the six T-injection instructions. On BB18, Bayes accuracy reaches 77.21\%, compared with 55.73\% for strict unique identification. The corresponding accuracies on BB72 are 97.85\% and 95.74\%. These results follow the same trends as logical Pauli measurement identification. Bayes accuracy exceeds strict unique identification because different candidates can produce the same classical measurement outcomes. The substantially higher accuracy on BB72 also reflects less overlap among candidate measurement outcomes than on BB18.

\subsubsection{Identification Under Noise}
Figure~\ref{fig:eval_t}(b) and (c) present scalable T-injection identification under circuit-level noise using the initial and final BB stabilizer measurement outcomes together with the auxiliary edge-\(Z\) readouts. On BB18, accuracy using the BB stabilizer and auxiliary edge-\(Z\) outcomes decreases from 77.34\% at LOW to 76.45\% at NOMINAL and 72.75\% at HIGH. On BB72, identification accuracy reaches 98.14\%, 98.24\%, and 95.31\% under LOW, NOMINAL, and HIGH noise, respectively. The LOW and NOMINAL results on both codes remain close to their ideal Bayes accuracies, while HIGH noise causes a moderate reduction. Notably, BB72 maintains identification accuracy above 95\% across all three noise profiles.
The lower accuracy on BB18 is associated with particular Pauli axes. Identification accuracy for \(YI\) and \(IY\) remains between 48.54\% and 57.31\% across the three noise profiles, with \(YI\) frequently identified as \(ZI\) and \(IY\) as \(IX\). In contrast, all six candidate axes on BB72 maintain identification accuracy above 93\% even under HIGH noise. This suggests that the lower accuracy on BB18 is associated with persistent difficulty in distinguishing certain Pauli axes, rather than circuit noise alone. We further examine the limits of distinguishing different logical operations from classical measurement outcomes in Section~\ref{subsec:identifiability}

\subsection{Logical-Circuit Identification}
\label{subsec:eval-circuits}
Finally, we evaluate logical-circuit identification. We first consider automorphism-based logical circuits consisting of automorphisms interleaved with SECs, and include DJ as a separately compiled workload. We then evaluate a logical circuit combining automorphisms, T injection, and adaptive logical Pauli measurement.

\subsubsection{Automorphism-Based Logical-Circuit Identification}
Figure~\ref{fig:eval_circuit_auto} presents the baseline scalable identification results for BB18 and BB72 under three noise profiles with PASSIVE and MARKER as defined in Section~\ref{sec:experimental-setup}. The AB, GHZ, BV, and HSP workloads contain automorphisms and SECs, while DJ uses directly compiled logical oracle circuits on BB18. Logical-circuit identification is the primary recovery target for all workloads, while net logical-action identification is reported separately for AB and GHZ, where different ordered circuits can produce the same final logical action.

With MARKER, logical-circuit identification accuracy on BB18 reaches 99.80\% to 100\% at LOW noise, 96.58\% to 99.02\% at NOMINAL, and 90.14\% to 94.34\% at HIGH across AB, GHZ, BV, and HSP. BB72 achieves similarly high accuracy of 99.71\%–100\% across these workloads at LOW noise. These results show that scalable inference can accurately recover logical circuits when a known Pauli error provides distinguishing syndrome information.

Similar to the exact automorphism inference results in Section~\ref{subsec:eval-auto}, scalable identification shows different trends under PASSIVE and MARKER as the error rate increases. This difference is most evident in AB on BB18, where PASSIVE logical-circuit identification increases from 6.25\% at LOW to 78.52\% at HIGH, while MARKER accuracy decreases from 100\% to 94.34\%. The shorter GHZ, BV, and HSP circuits show smaller PASSIVE increases, reaching 33.50\% to 37.01\% at HIGH. The longer AB circuit provides more opportunities for naturally occurring errors to be transported and captured in subsequent measured syndromes, which is consistent with its larger PASSIVE improvement. On BB72, identification becomes substantially more sensitive to noise even though it shares similar trends as BB18. At NOMINAL, MARKER accuracy decreases to 58.98\% for AB and 81.64\%–83.89\% for GHZ, BV, and HSP. At HIGH noise, it falls to 0.39\% for AB and 11.82\% - 14.45\% for the other workloads, with similarly low accuracy under PASSIVE. This sharp degradation at HIGH noise motivates the configuration analysis in Section~\ref{subsec:scalable_approximation}, which evaluates broader configurations on the same shots.

The recovery target dependence on both codes also follows the exact automorphism identification results in Section~\ref{subsec:eval-auto}. Different logical circuits can share the same net logical action, allowing the net logical action to be identified even when the logical circuit remains unresolved. For example, at LOW noise under PASSIVE, AB identification accuracy increases from 6.25\% for the ordered circuit to 41.50\% for the net logical action on BB18, and from 20.90\% to 50.59\% on BB72.

The separately compiled DJ workload on BB18 behaves differently from the other workloads. Logical-circuit identification remains close to the 25\% reference across LOW, NOMINAL, and HIGH noise, and MARKER does not provide the large improvement observed for AB, GHZ, BV, and HSP. Section~\ref{subsubsec:syndrome_identifiability} explains the detailed reason behind this behavior.

\subsubsection{Logical-Circuit Identification with Three Instruction Types}
We evaluate a non-Clifford Pauli-rotation circuit combining all three instructions as defined in Section~\ref{sec:experimental-setup}. The combined scalable identification method in Section~\ref{subsec:method-circuit} achieves an average logical-circuit identification accuracy of 78.32\% across eight candidates under LOW circuit-level noise with MARKER. Identification uses the joint candidate likelihood of the complete circuit rather than combining independent identifications of its constituent instructions.

Figure~\ref{fig:eval_circuit_mix} shows identification accuracies across candidate circuits. The four circuits with \(P=ZI\) achieve 92.19\%–96.88\% accuracy, averaging 94.92\%, while those with \(P=YI\) achieve 54.69\%–66.41\%, averaging 61.72\%. To understand this difference, we examine the incorrect identifications. Among these incorrect results, 85.07\% misidentify \(P=YI\) as \(P=ZI\) while correctly identifying automorphisms and logical Pauli measurements. Overall, our logical-circuit identification achieves high accuracy, with occasional misidentification on the T-injection axis.

\section{Identifiability Discussion: When Can Logical Operations Be Identified?}
\label{sec:limit}

% \Shifan{Indentifiability discussion: why can't we identify all the errors?}
\begin{figure*}[t]
    \centering
    \includegraphics[width=\textwidth]{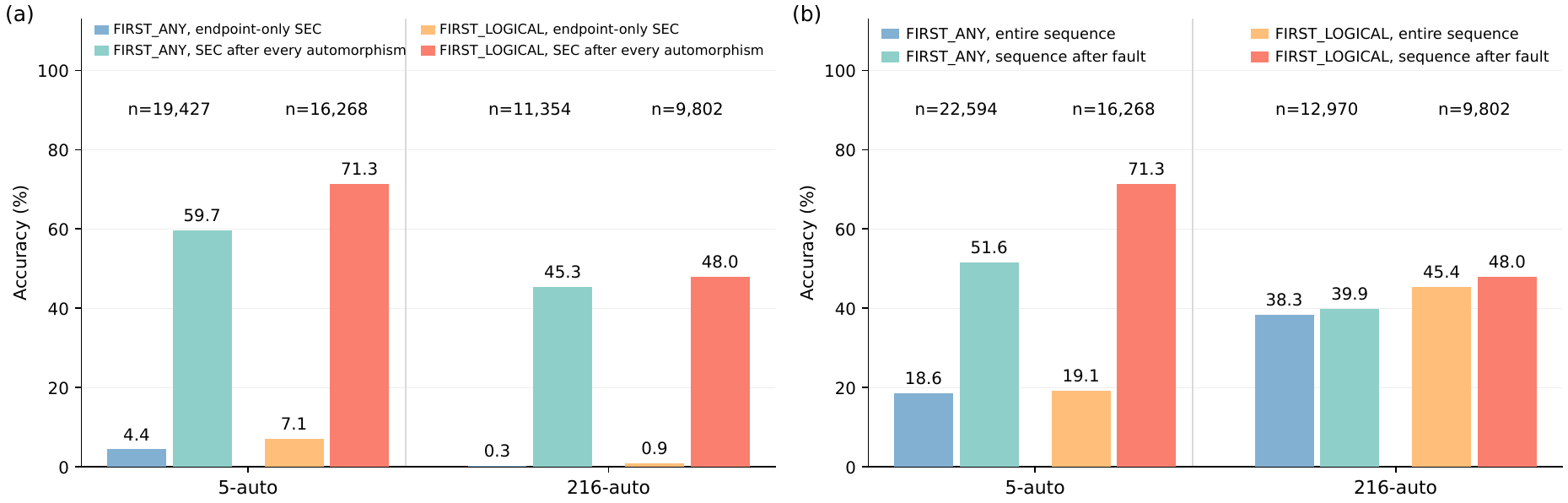}
    \caption{PASSIVE analysis of syndrome-extraction schedule and fault timing for the BB18 Exact-Sum ordered automorphism sequence experiments. Results are pooled across the noise models and rates evaluated in Section~\ref{subsec:eval-robustness}. (a) Identification accuracy for the ordered automorphism sequence executed after the fault under syndrome extraction after every automorphism and endpoint-only syndrome extraction. (b) Identification accuracy for the entire ordered automorphism sequence and for the ordered automorphism sequence executed after the fault, using syndrome extraction after every automorphism. Each \(n\) is the number of PASSIVE shots included after applying the stated fault alignment and requiring at least one automorphism after the fault. FIRST\_ANY and FIRST\_LOGICAL can therefore have different values of \(n\).}
    \label{fig:limit_passive_auto}
\end{figure*}

\begin{figure*}[t]
    \centering
    \includegraphics[width=\textwidth]{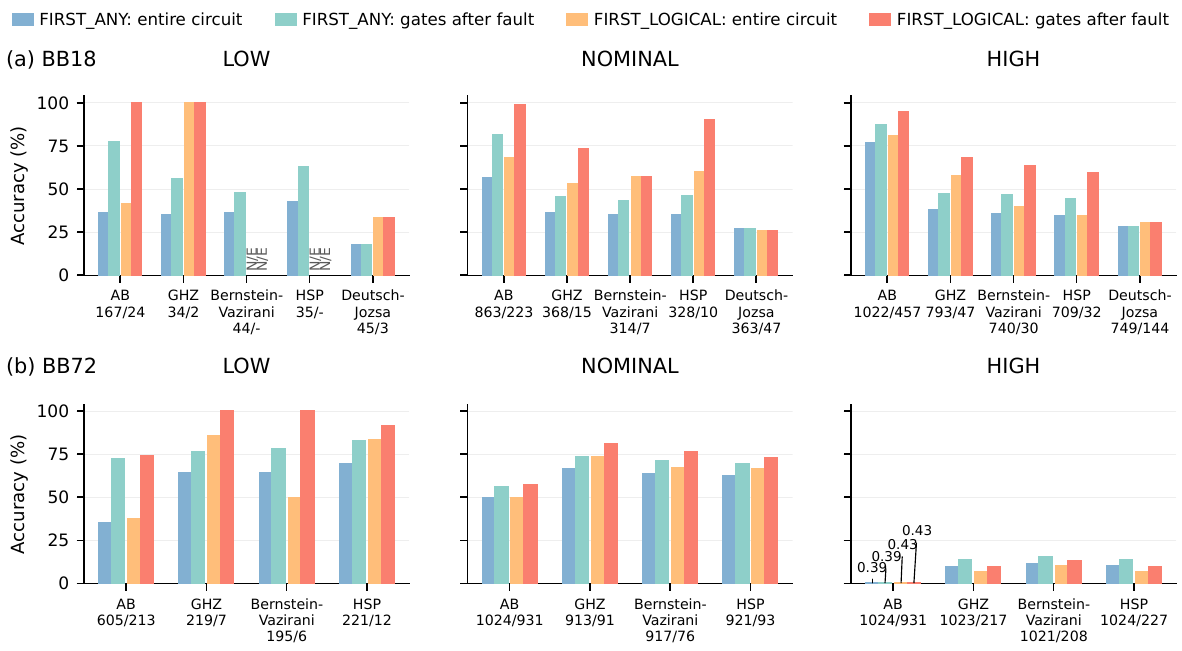}
    \caption{PASSIVE fault-timing analysis for the logical-circuit workloads using the baseline scalable inference configuration across (a) BB18 and (b) BB72 codes tested in Section~\ref{subsec:eval-circuits}. Bars compare identification of the entire logical circuit with identification of the ordered logical gates executed strictly after the fault on the same shots. Labels \(n_A/n_L\) give the numbers of shots included under FIRST\_ANY and FIRST\_LOGICAL, respectively, after requiring at least one logical gate after the fault. A dash indicates that no shot satisfies this condition, so the corresponding result is not estimable.}
    \label{fig:limit_passive_circuit}
\end{figure*}

\begin{figure*}[t]
    \centering
    \includegraphics[width=\textwidth]{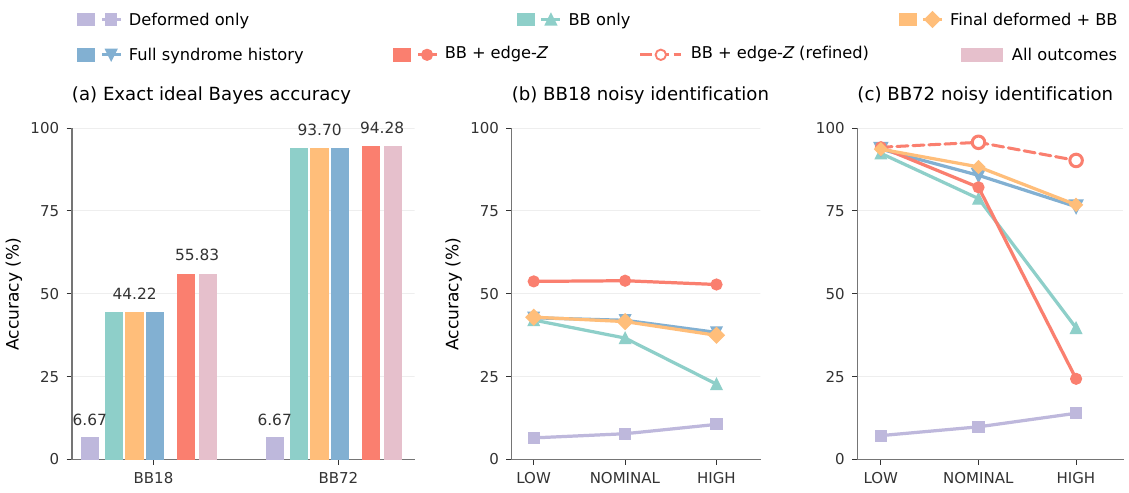}
    \caption{Logical Pauli measurement identification using different classical measurement outcomes and scalable inference refinement. (a) Exact ideal Bayes accuracy on BB18 and BB72. (b) BB18 and (c) BB72 identification accuracy under LOW, NOMINAL, and HIGH circuit-level noise. Deformed only uses the deformed stabilizer measurements from both gauging rounds. BB only uses the initial and final BB stabilizer measurements. Final deformed + BB uses the final deformed and BB stabilizer measurements. Full syndrome history includes the initial and final BB stabilizer measurements and both rounds of deformed stabilizer measurements. BB + edge-\(Z\) includes the initial and final BB stabilizer measurements and auxiliary edge-\(Z\) readouts as in Section~\ref{subsec:eval-pm}. Refined denotes the modified intermediate probability-state selection on BB72 NOMINAL and HIGH evaluated in Section~\ref{subsec:scalable_approximation}. All outcomes includes the complete classical measurement outcomes.}
    \label{fig:limit_pm}
\end{figure*}

\begin{figure*}[t]
    \centering
    \includegraphics[width=\textwidth]{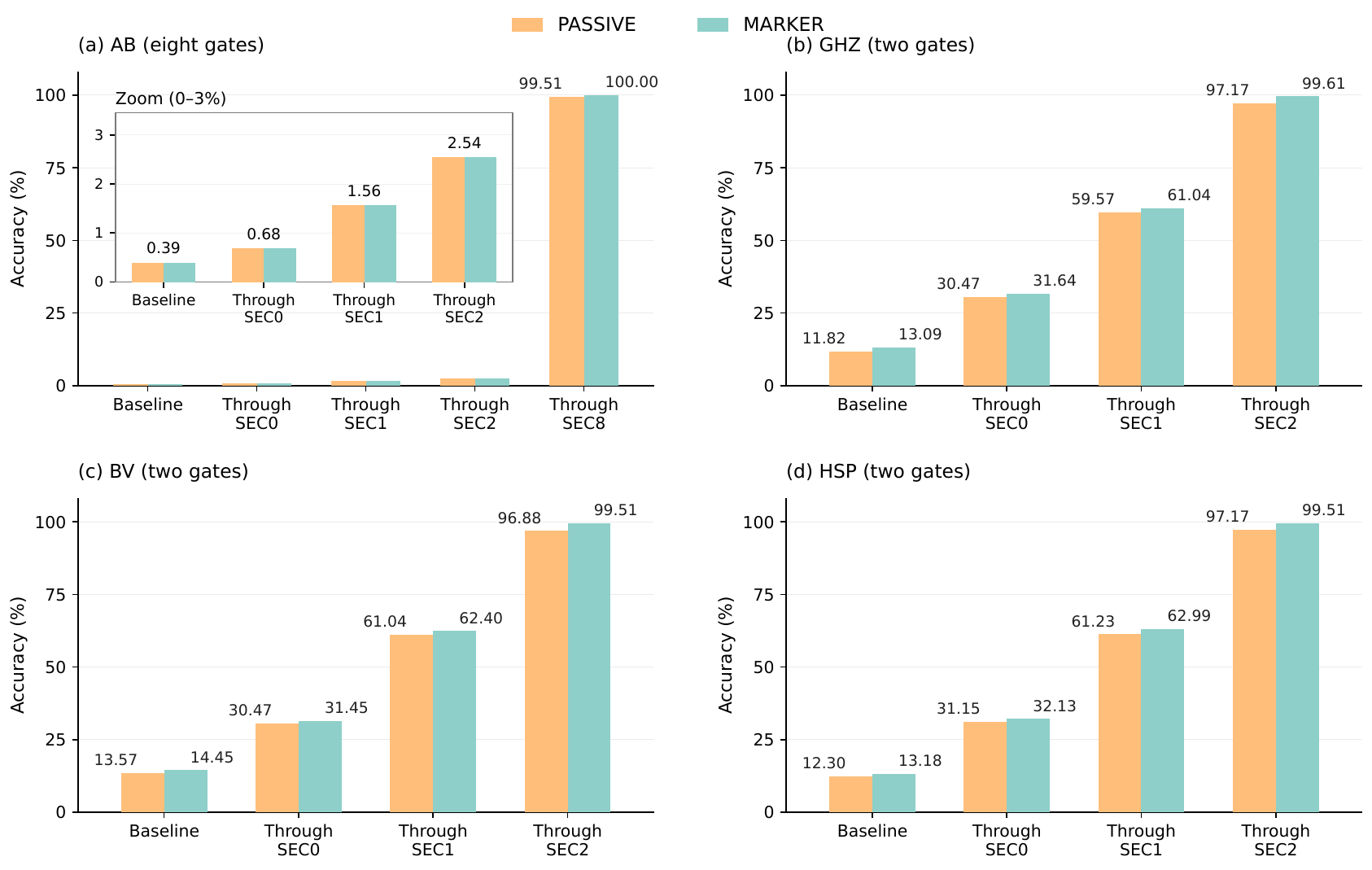}
    \caption{Scalable inference refinement for logical-circuit identification on BB72 HIGH across (a) AB, (b) GHZ, (c) BV, and (d) HSP. Bars report logical-circuit identification accuracy under PASSIVE and MARKER on the same shots used for the baseline results in Section~\ref{subsec:eval-circuits}. The scalable inference configurations are defined in Section~\ref{sec:experimental-setup}. “Through SECx” expands the fault-pattern calculation through the indicated SEC boundary, while subsequent circuit blocks retain the baseline configuration.}
    \label{fig:limit_scale_auto}
\end{figure*}

The preceding results show that identification can fail for different reasons. First, the available classical measurement outcomes may not distinguish the requested recovery targets. Second, for automorphism identification, distinguishing information from error transport depends on the occurrence and timing of informative data errors. Third, scalable inference may omit fault patterns or intermediate states needed to use information present in the classical measurement outcomes. We examine these three limitations and corresponding refinements below.

\subsection{Identifiability}
\label{subsec:identifiability}
The classical measurement outcomes do not always provide sufficient information to identify the executed logical operation. For automorphism identification, different candidates may produce the same measured syndrome histories. For logical Pauli measurements and T injections, the information available for identification also depends on which classical measurement outcomes are used. Some outcomes provide no distinguishing information, while others can distinguish candidates but, as in automorphism identification, may not always uniquely identify the executed instruction. Since T-injection identification relies on the same gauging and ungauging correlations as logical Pauli measurement identification, we focus on automorphism and logical Pauli measurement identification in this section.

\subsubsection{Syndrome Identifiability}
\label{subsubsec:syndrome_identifiability}
For some operations, the syndrome histories provide no distinguishing information. If two candidate sequences associated with different recovery targets assign the same likelihood to every possible syndrome history under the specified SEC schedule and noise model, no method using only that syndrome history can identify the recovery target correctly for both candidates with certainty.

The exact experiments in Section~\ref{subsec:eval-auto} already show an example of this limitation. Among the 200 sampled ordered automorphism sequences, only 193 distinct ideal MARKER syndrome histories are obtained even when syndrome extraction is performed after every automorphism.

DJ provides another logical-circuit example in Section~\ref{subsec:eval-circuits}. The BB18 DJ workload contains four encoded quadratic-phase oracle circuits, one constant and three balanced. With a uniform prior, an uninformative four-way decision gives a 25\% reference accuracy. DJ identification remains close to this value across three noise profiles and PASSIVE/MARKER conditions. The DJ result follows from how its encoded \(Z\) and \(CZ\) operations affect the stabilizers measured by the SECs. At the SEC boundaries used in this experiment, these operations leave the measured stabilizers unchanged. Therefore, for the same incoming data error, the four DJ circuits produce the same ideal syndrome change after the operation. Under the noisy circuit model, however, faults within the compiled circuits can still produce different measured syndrome histories for different candidates. Their noisy candidate likelihoods therefore need not be exactly equal, which is consistent with the identification accuracy being close to the 25\% reference. Thus, DJ shows that different logical circuits can remain difficult to distinguish when they do not produce different measured syndromes at the SEC boundaries used for identification.

\subsubsection{Logical Pauli Measurement Identifiability}
\label{subsubsec:pm_identifiability}
We next examine how different classical measurement outcomes affect logical Pauli measurement identification, first in the ideal setting and then under circuit noise. We particularly investigate whether stabilizer syndrome measurements alone can support identification, with similar implications for T-injection identification.

Figure~\ref{fig:limit_pm}(a) compares the exact ideal Bayes accuracy using different classical measurement outcomes. Although different logical Pauli measurements use different deformed stabilizer checks, all 15 candidates produce zero deformed stabilizer outcomes in both measurement rounds under ideal execution.
Deformed only therefore achieves 6.67\% accuracy, corresponding to uniform guessing among the 15 candidates. In contrast, using only the BB syndromes measured before and after the logical Pauli measurement (BB only) achieves 44.22\% on BB18 and 93.70\% on BB72. Replacing the initial BB syndrome with the final deformed syndrome (Final deformed + BB), or retaining both BB syndromes and both rounds of deformed stabilizer measurements (Full syndrome history), gives the same ideal Bayes accuracies. Including auxiliary edge-Z outcomes with final BB syndromes (BB + edge-\textbf{Z}) as in Section~\ref{subsec:eval-pm} increases these values to 55.83\% and 94.28\%, respectively. Combining all BB syndromes and classical measurement outcomes during logical Pauli measurement provides no further improvement. However, even when all classical measurement outcomes are retained, a single shot may not uniquely identify the measured logical Pauli operator. Although all 15 candidates produce different correlations among the outcome bits, the same classical measurement outcomes may still occur under multiple candidates. This overlap is substantially greater on BB18 than on BB72, helping explain the difference in their ideal identification accuracy.

Figure~\ref{fig:limit_pm}(b) and (c) compare scalable identification using different classical measurement outcomes under circuit noise.
Identification using deformed stabilizer outcomes increases with the noise rate, from 6.47\% to 10.55\% on BB18 and from 7.16\% to 13.91\% on BB72. At higher noise rates, these approximate accuracies modestly exceed the 6.67\% ideal Bayes accuracy because faults can introduce operation correlated information into the deformed stabilizer outcomes. However, the identification remains limited. In contrast, BB only achieves 42.05\% on BB18 and 92.38\% on BB72 under LOW noise, close to their ideal Bayes accuracies, but decreases to 22.70\% and 39.67\%, respectively, under HIGH noise. This degradation is consistent with the lack of intermediate syndrome measurements during the logical Pauli measurement, which makes it more difficult to distinguish the effects of faults occurring within the procedure as the noise rate increases. Final deformed + BB retains a syndrome measurement within the logical Pauli measurement and shows less degradation, achieving 37.44\% on BB18 and 76.85\% on BB72 under HIGH noise. Full syndrome history follows a similar trend but does not consistently improve scalable identification accuracy. These results show that selecting the syndrome measurements for identification can be more important than simply retaining more syndrome outcomes under circuit noise.

Including auxiliary edge-Z outcomes provides a further improvement on BB18. BB + edge-Z maintains identification accuracy between 52.78\% and 53.92\% across all three noise profiles, close to its ideal Bayes accuracy of 55.83\%. On BB72, BB + edge-$Z$ achieves 94.22\% under LOW noise, close to the 93.64\% obtained with Final deformed + BB, but decreases to 82.12\% and 24.31\% under NOMINAL and HIGH noise, respectively. 
We examine the BB72 accuracy degradation and its improvement through scalable inference refinement in Section~\ref{subsec:scalable_approximation}.

These results show that stabilizer syndrome measurements alone are sufficient to support logical Pauli measurement identification. Since T-injection identification builds on the logical Pauli measurement identification method, this input reduction could also be extended to T injection, allowing all three logical instructions to be identified using only syndromes.

\subsection{Fault Timing and Availability of Distinguishing Information}
\label{subsec:fault_timing_and_sec}
Section~\ref{subsec:eval-auto} shows that a data error is useful for automorphism identification only when different operations transport it differently and the SEC schedule records the resulting syndrome changes. In contrast, logical Pauli measurement and T-injection identification can use candidate-dependent correlations among classical measurement outcomes even without stochastic faults. We therefore focus on how fault timing affects automorphism identification and automorphism-based logical-circuit identification under PASSIVE, where distinguishing syndrome information from error transport depends on when a data error occurs.

In PASSIVE, errors are sampled naturally from the declared stochastic noise model rather than deliberately inserted. The resulting shots therefore include cases with no data error and cases in which a data error appears only after part of the circuit has already executed. Only data errors that are subsequently transported by later operations provide information about those later operations. We therefore use the FIRST\_ANY and FIRST\_LOGICAL, as defined in Section~\ref{sec:experimental-setup}, to analyze identification accuracy after the first fault. Fault locations are used only for this retrospective analysis and are not supplied to the likelihood calculation.

Figure~\ref{fig:limit_passive_auto}(a) compares identification under different syndrome-extraction schedules. Under FIRST\_LOGICAL, identification accuracy for the ordered automorphism sequence executed after the fault reaches 71.3\% with syndrome extraction after every automorphism but only 7.1\% with endpoint-only syndrome extraction for the five-automorphism experiment. The broader 216-automorphism experiment gives 48.0\% and 0.9\%, respectively. FIRST\_ANY shows the same qualitative difference. Fault occurrence alone therefore does not guarantee identification. The syndrome history must also capture that distinguishing error transport by subsequent candidate operations. The measurement-error-only results in Section~\ref{subsec:eval-robustness} provide the complementary case. Measurement errors change the measured syndromes but do not place a Pauli error on the data for later operations to transport. Figure~\ref{fig:limit_passive_auto}(b) further compares identification of the entire ordered automorphism sequence with identification of the ordered automorphism sequence executed after the fault. On the same shots, restricting the recovery target to the automorphisms executed strictly after the fault increases accuracy from 19.1\% to 71.3\% for the five-automorphism experiment. This result shows that a late data error can leave earlier automorphisms unresolved while still providing information about later automorphisms. The smaller change for the broader set, from 45.4\% to 48.0\%, shows that shortening the target does not by itself guarantee high recovery.

Figure~\ref{fig:limit_passive_circuit} compares these recovery targets in the automorphism-based logical-circuit experiments in Section~\ref{subsec:eval-circuits}. For BB18 AB at NOMINAL noise, FIRST\_LOGICAL identification increases from 68.16\% for the entire logical circuit to 99.10\% for the logical gates executed strictly after the fault. For BB72 AB at LOW noise, the corresponding values are 37.56\% and 74.18\%. Several GHZ, BV, and HSP settings show similar trends. DJ is limited by syndrome identifiability as discussed in Section~\ref{subsubsec:syndrome_identifiability}. BB72 HIGH is addressed separately in Section~\ref{subsec:scalable_approximation}.

\subsection{Scalable Inference Approximation and Refinement}
\label{subsec:scalable_approximation}
The final limitation arises from the scalable inference approximation. Larger codes and longer circuits contain more fault locations, so the same noise rates increase the expected number of faults and the contribution from multi-fault events. A scalable configuration that considers too few fault patterns or retains an insufficient set of intermediate states may therefore omit contributions needed to distinguish recovery targets, even when the classical measurement outcomes contain distinguishing information. This limitation is particularly apparent in the BB72 logical Pauli measurement results under NOMINAL and HIGH noise in Section~\ref{subsec:eval-pm} and the BB72 logical-circuit results under HIGH noise in Section~\ref{subsec:eval-circuits}. We therefore evaluate different refinements for these two identification tasks.

\subsubsection{Logical Pauli Measurement Inference Refinement}
Figure~\ref{fig:limit_pm}(c) evaluates refinement of intermediate probability-state selection for logical Pauli measurement identification on BB72 under NOMINAL and HIGH. As described in Section~\ref{subsec:method-pm}, the original configuration prioritizes intermediate probability states using their accumulated probabilities and conservative upper bounds on the probability of producing the required parity-check violation in the remaining circuit stages. The refinement instead combines projected probabilities across multiple parity-check subsets to select states more likely to produce the required final violation. This changes how states are selected without increasing the violation-state budget \(B_V=4096\) or changing the likelihood-bounding rules.

Compared with baseline configuration in Section~\ref{subsec:eval-pm}, identification accuracy increases from 82.12\% to 95.70\% under NOMINAL noise and from 24.31\% to 90.23\% under HIGH noise with refinement. Identification availability also improves from 88.28\% and 35.25\%, respectively, to 100\% for both noise profiles. These results demonstrate that more effective intermediate probability-state selection can substantially improve identification accuracy under the same state budget. The improvement on the same shots shows that low baseline identification accuracy can arise from limitations of scalable inference rather than insufficient distinguishing information in the classical measurement outcomes alone.

\subsubsection{Logical-Circuit Inference Refinement}
We next examine whether expanding the fault-pattern budget improves logical-circuit identification. Figure~\ref{fig:limit_scale_auto} illustrates the expanded scalable configuration defined in Section~\ref{sec:experimental-setup} on the same BB72 HIGH shots. We apply the expanded configuration to progressively longer portions of each circuit, while retaining the baseline configuration for subsequent circuit blocks.

Figures~\ref{fig:limit_scale_auto}(b)--(d) show the corresponding results for GHZ, BV, and HSP. Each of them contains only two logical gates.  Extending the broader calculation through SEC2 covers the full logical circuit and raises accuracy from approximately 12\%–14\% at baseline to 97\%–99.6\% on the same shots. Since the physical circuits and syndrome histories are unchanged, this increase directly shows that the baseline scalable configuration, rather than the available syndrome history alone, was the dominant limitation in these settings. These results emphasize the importance of sufficient fault-pattern coverage under the fixed hidden boundary-state budget.

For the eight-gate AB circuit, partial expansion initially produces much smaller improvements. Logical-circuit identification accuracy increases from 0.39\% at baseline to 0.68\% with the broader calculation through SEC0, 1.56\% through SEC1, and 2.54\% through SEC2 under both PASSIVE and MARKER. However, expanding the calculation through SEC8 covers the entire eight-gate circuit and raises accuracy to 99.51\% under PASSIVE and 100\% under MARKER. These results show that expanding only the early circuit blocks does not prevent the loss of candidate likelihoods in subsequent blocks, whereas full-circuit expansion substantially improves identification.

\section{Related Work}
\label{sec:related_work}

Recent studies show that information exposed during quantum execution can reveal properties of the hidden circuit. One reconstructs gate-level circuits from quantum controller power traces by first recovering control-pulse information and then inferring the executed circuit~\cite{erata2024quantum}. Another approach observes the space-time activity of surface-code patches during lattice surgery, reconstructs logical-qubit dataflow, and applies graph matching to identify subroutines or complete programs~\cite{trochatos2026traceq}. These approaches infer computation from control or resource-activity traces. In contrast, our method uses syndrome histories and optionally ancillary measurement outcomes produced during logical-circuit execution, whose dependence on the executed operation arises from error transport and operation-dependent measurement correlations.

Error-related measurements have also been proposed as fingerprints for hidden properties other than the executed logical operation. Error-syndrome statistics have been  to identify and authenticate quantum hardware backends~\cite{mutolo2025quantum}, while localized noise fingerprints have been used to detect circuit tampering~\cite{macneil2025authenticating}. These studies show that QEC and noise information can expose metadata beyond its conventional role in error correction.

The closest related work infers hidden logical operations from gate-dependent fingerprints in surface-code spacetime syndromes~\cite{shukla2026anticipating}. It formulates operation inference by marginalizing physical error patterns and demonstrates that different fault-tolerant gates can produce distinguishable syndrome statistics. In its evaluation, however, the full marginalized likelihood is not computed for circuit reconstruction. Single-gate inference instead uses an approximate classifier, and higher-level circuit structure is subsequently used to recover program information. The work therefore establishes the existence and potential leakage of operation-dependent syndrome fingerprints, but it does not provide a general inference procedure for noisy multi-operation histories.
Our work develops this inference direction for BB logical operations. We calculate candidate likelihoods using syndrome histories and relevant ancillary measurement outcomes, account for unknown faults and dependencies across logical instructions, and develop exact and scalable identification methods for logical instructions and logical circuits.

\section{Discussion and Future Work}
\label{sec:discussion}

Our results show that syndrome histories and other classical measurement outcomes can contain information about executed logical operations, in addition to the error information normally used for decoding. This information can arise from error transport under automorphisms or operation-dependent correlations between stabilizer and ancillary measurements. These findings suggest potential applications in execution diagnostics and raise questions about information exposure. For a predefined set of logical instructions, identification from classical measurement outcomes could provide an independent means of checking which operation was executed, without relying solely on the intended circuit description. Conversely, access to these outcomes may reveal information about the executed logical operations, motivating further study of the confidentiality of classical measurement interfaces. However, our experiments do not establish the feasibility of real-time identification or quantify its hardware overhead and security implications.

Our framework could also be extended to other fault-tolerant logical operations, such as transversal gates and operations implemented through code deformation, by developing candidate likelihood models that capture their physical error propagation and classical measurement outcomes. Such extensions would broaden the supported logical instruction set without changing the identification objective. Future work can study longer adaptive logical circuits, additional qLDPC code families, and more efficient inference methods. In particular, reusing intermediate computations across candidate operations and circuit segments, together with more selective fault-pattern evaluation and adaptive refinement, could reduce redundant computation while maintaining identification accuracy. Finally, extending the current closed candidate setting to unknown-operation rejection, model mismatch, and hardware measurement records would help determine how broadly logical-operation identification can be used in practical quantum systems.

\section{Conclusion}
\label{sec:conclusion}

We developed exact identification methods and scalable approximations for logical-operation identification in qLDPC codes. The methods exploit error transport under code automorphisms and candidate-dependent measurement correlations in logical Pauli measurements and T injections, while accounting for unknown faults and measurement randomness. Experiments on \([[18,4,2]]\) and \([[72,12,6]]\) demonstrate logical-instruction and logical-circuit identification under different noise conditions. We further show that stabilizer syndrome histories alone can identify gauging-based logical Pauli measurements, extending syndrome-based identification beyond automorphisms without requiring auxiliary readouts for inference. Together, these results show that syndrome histories can provide information about executed logical operations and circuits beyond their conventional role in error decoding.

\section*{Acknowledgements}

%%%%%%% -- PAPER CONTENT ENDS -- %%%%%%%%

%%%%%%%%% -- BIB STYLE AND FILE -- %%%%%%%%
\bibliographystyle{IEEEtranS}
\bibliography{refs}
%%%%%%%%%%%%%%%%%%%%%%%%%%%%%%%%%%%%
% \clearpage
% \appendices
% \input{sections/appendix_exact_auto}

\end{document}